\documentclass[12pt]{article}

\usepackage[margin=0.95 in]{geometry}
\usepackage{amsmath} 
\usepackage{setspace}
\usepackage{slashed}
\usepackage{comment} 
\usepackage{dsfont}
\usepackage{amssymb,amsfonts}
\usepackage[all]{xy}
\usepackage{graphicx}
\usepackage{relsize}
\usepackage{amsmath}
\usepackage{amssymb}
\usepackage{float}
\usepackage{array}
\usepackage{tikz}
\usepackage{mathtools}
\usepackage{mathrsfs} 
\usepackage{cite}
\usepackage{tikz}
\usepackage{array, multirow}  

\numberwithin{equation}{section}

\newcommand{\be}{\begin{equation}}
\newcommand{\ee}{\end{equation}}

 \newcommand{\Tr}{{\text{Tr}}}
  
\def\bea{\begin{eqnarray}}
\def\eea{\end{eqnarray}}
\newcommand{\R}{\mathrm{R}}
\newcommand{\NS}{\mathrm{NS}}
\allowdisplaybreaks

\numberwithin{equation}{section}
\numberwithin{table}{section}

\usepackage[colorlinks=true, citecolor=black, linkcolor=black, filecolor=black,urlcolor=blue]{hyperref}
\usepackage{amsmath}
\usepackage{amsfonts}
\usepackage{amssymb}
\usepackage{graphicx}
\usepackage{grffile}
\usepackage{subfig}
\usepackage{hyperref}
\usepackage{multirow}
\usepackage{physics}
\usepackage{comment}

\usepackage{braket}
\usepackage{xcolor}
\usepackage{cite}
\usepackage{dsfont}
\usepackage{float}
\usepackage{ulem}
\usepackage{breqn}
\usepackage{tikz}
\def\be{\begin{equation}}
\def\ee{\end{equation}}
\def\bea{\begin{eqnarray}}
\def\eea{\end{eqnarray}}

\newcommand{\Dc}{\mathcal{D}}

\tikzset{
  pics/torus/.style n args={3}{
    code = {
      \providecolor{pgffillcolor}{rgb}{1,1,1}
      \begin{scope}[
          yscale=cos(#3),
          outer torus/.style = {draw,line width/.expanded={\the\dimexpr2\pgflinewidth+#2*2},line join=round},
          inner torus/.style = {draw=pgffillcolor,line width={#2*2}}
        ]
        \draw[outer torus] circle(#1);\draw[inner torus] circle(#1);
        \draw[outer torus] (180:#1) arc (180:360:#1);\draw[inner torus,line cap=round] (180:#1) arc (180:360:#1);
      \end{scope}
    }
  }
}

\begin{document}

\hypersetup{pageanchor=false}
\begin{titlepage}
\vbox{\halign{#\hfil    \cr}}  

\begin{center}

{\Large \bf 
Modular properties of correlators in\\ generalised Gibbs ensembles \\[2mm] 
}

\vspace*{5mm}

{\large 
 Tanmoy Sengupta  \\
}

\vspace{.5cm}

{Chennai Mathematical Institute, \\H1, SIPCOT IT Park, 
Siruseri, Kelambakkam, 603103, India.\\}

 \vskip 0.8cm
	{\small
		E-mail:\texttt{tanmoys@cmi.ac.in } \\
  
	}
\vspace*{0.8cm}

\end{center}

\begin{abstract} 
We study the modular properties of torus correlators in a
generalised Gibbs ensemble (GGE) of a two-dimensional
conformal field theory. For the free fermion with KdV
charges, we compute the exact two-point function and derive
its modular $S$ transformation as a formal power series in
the chemical potential. The transformed correlator admits
a factorised description: an effective charge deforms the
ensemble, while a local dressing operator acts on the
external fields. Both are determined by nested operator
product expansions involving the deforming current.
Motivated by this result, we conjecture an analogous
transformation law for $n$-point functions in rational
chiral conformal field theories. Using Zhu's recursion,
we establish the conjecture at the lowest explicit power
of the modular parameter to all orders in the chemical
potential, thereby determining the dressing operator.
We also verify the full conjectured transformation through
cubic order in the chemical potential, retaining all
explicit powers of the modular parameter. These results
extend the nested-OPE description of modular transformations
of GGE partition functions to local correlation functions.

\end{abstract}

\end{titlepage}

\hypersetup{pageanchor=true}
\setcounter{tocdepth}{2}

{\begin{spacing}{0.1}
\tableofcontents
\end{spacing}

\newpage
\section{Introduction}
\label{sec:intro}

Two-dimensional conformal field theories (CFTs) \cite{Belavin:1984vu} are a cornerstone of string theory, critical phenomena and mathematical physics \cite{Friedan:1980jm, ZamolodchikovsBook, DiFrancesco:1997nk}. One of their most remarkable features is that they possess infinitely many commuting conserved charges. In a CFT with Virasoro symmetry, these are operators in the universal enveloping algebra of the Virasoro algebra. They are called quantum KdV charges, because at large central charge they reduce to the conserved charges of the classical Korteweg--de Vries (KdV) hierarchy \cite{Sasaki:1987mm, Eguchi:1989hs, Kupershmidt:1989bf, Bazhanov:1994ft, Bazhanov:1996dr, Bazhanov:1996aq, Bazhanov:1998dq}. 

A system with conserved charges beyond the Hamiltonian should be described in equilibrium by a generalised Gibbs ensemble (GGE) rather than by the ordinary Gibbs ensemble \cite{Rigol:2007, Vidmar:2016}. One turns on a chemical potential $\mu_i$ for every conserved charge $I_i$, and the density matrix is
\begin{equation}
\label{eq:rhoGGE}
\rho_{\text{GGE}}\propto\;\exp\Big(-\beta\big(L_0-\tfrac{c}{24}\big)+\sum_{i}\mu_{i}I_{i}\Big)\,,
\end{equation}
where $\beta$ is the inverse temperature. Given that an integrable system has infinitely many conserved charges, the question of whether it relaxes to a Gibbs ensemble after a quench has been a central question in many-body physics. Experiments ranging from the quantum Newton's cradle \cite{Kinoshita:2006} to nonlinear optical fibres \cite{Bastianello:2025} point to GGEs as the correct description of late-time relaxation in classical and quantum integrable systems. In two-dimensional CFTs, ensembles built from the KdV charges have been studied in \cite{Maloney:2018hdg, Dymarsky:2018lhf}, and more recently through the ODE/IM correspondence \cite{Dorey:1998pt, Bazhanov:1998wj, Ashok:2026mfy}.

A CFT on a torus has modular symmetry, which has been used to derive powerful results such as the Cardy formula for the asymptotic density of states~\cite{Cardy:1986ie}. The characters of the Virasoro minimal models can similarly be shown to transform as vector-valued modular forms. It is therefore natural to ask how the generalised characters, obtained by taking the trace of $\rho_{\text{GGE}}$, transform. When chemical potentials for higher-spin charges are turned on, ordinary modular covariance is lost, and the modular transformation of the partition function becomes non-trivial. This problem was first solved in a special case in~\cite{Dijkgraaf:1996iy}, and there has been considerable recent progress towards a solution of this problem. For ensembles with KdV charges, the modular properties of such generalised characters were studied in~\cite{Downing:2021mfw, Downing:2024nfb, Karimi:2026dnr, Downing:2023lop, Downing:2023lnp}, and for the Boussinesq charges of symplectic free fermions in~\cite{Karimi:2026tvh}. For generic central charge, the modular properties in the presence of the first non-trivial Boussinesq charge, the zero mode of the spin-three current, were studied as an asymptotic series in the chemical potential. By making use of the modular properties of traces of zero modes~\cite{Iles:2013jha, Iles:2014gra}, a conjecture was proposed in~\cite{Downing:2025huv} as to the modular-transformed generalised character. This conjecture was derived, and the result extended to arbitrary chiral deformations, in~\cite{Ashok:2026cyi}.

All of this concerns the generalised characters, which feed into the calculation of the partition function. A partition function encodes the spectrum of the theory weighted by the conserved charges, and hence the thermodynamics of the ensemble. An ensemble, however, is ultimately characterised by the expectation values of observables, and the GGE hypothesis itself is a statement about local observables: at late times, correlation functions of local operators in a quenched integrable system are reproduced by the GGE~\cite{Rigol:2007, Vidmar:2016}. In a CFT, the prototype of this statement is that correlation functions of local operators after a global quench relax to thermal ones, with an effective temperature fixed by the energy~\cite{Calabrese:2006rx}. In the presence of the KdV charges, the thermal ensemble should be replaced by a GGE. Correlation functions of local operators are therefore the natural objects by which GGEs are tested, and they are also the quantities accessible in experiments.

A second motivation comes from holography. At large central charge, a GGE with chemical potentials for higher-spin charges is dual to a higher-spin black hole, whose thermodynamics was reproduced from the partition function of a higher-spin ${\cal W}$-algebra in \cite{Gaberdiel:2012yb}. Probe fields propagating in such backgrounds are dual to correlation functions of local operators in the GGE. They were computed in the bulk using Wilson lines in \cite{Kraus:2012uf} and compared with CFT computations in \cite{Gaberdiel:2013jca}. Control over the modular properties of correlators in GGEs would extend such comparisons to all orders in the chemical potential at high temperature, since the modular transformation relates the high-temperature regime to the low-temperature regime, where the vacuum dominates.

There are additional motivations to study correlation functions in GGEs. In the presence of chemical potentials, the modular properties of torus correlators have been used to constrain structure constants~\cite{Das:2017vej}. For GGEs, the analogous statements require knowing how correlators transform. Moreover, the modular transformation relates the GGE to a description in the crossed channel in terms of a line defect~\cite{Downing:2023lnp}, so that such a transformation also gives correlation functions in the presence of line defects. Finally, two-point functions can be used to compute the Krylov complexity of local operators \cite{Parker:2018yvk, Dymarsky:2021bjq, Avdoshkin:2022xuw}.

In this paper, we initiate the study of how correlation functions of local fields in a GGE behave under modular transformations. We start with the free Majorana fermion deformed by KdV charges, where the correlators on the torus can be computed exactly. Using the modular properties of derivatives of Jacobi forms, we determine the modular transformation of the fermion two-point function to all orders in the chemical potential. The result has a simple structure. The transformed correlator is governed by the same effective current that controls the partition function \cite{Downing:2025huv, Ashok:2026cyi}, but the inserted fields are in addition dressed. The dressing is generated by the nested operator product expansions of the inserted fields with the currents that make up the effective current. Since this structure only involves OPE data, it makes sense in any chiral algebra, and we formulate a conjecture for correlators of chiral fields in a generic chiral algebra. To test it, we use Zhu's recursion relations \cite{zhu:1990, zhu:1996} to reduce the modular transformation at each order in the chemical potential to integrals of correlators over the $B$-cycle of the torus. We establish the conjecture at the lowest explicit power of the modular parameter, to all orders in the chemical potential, thereby determining the dressing operator. This part does not involve the zero modes of the effective current, which enter only at higher explicit powers of the modular parameter. With the dressing operator fixed, these terms provide an independent test of the conjecture, which we verify through order $\mu^3$, finding agreement.

The paper is organised as follows. In Section~\ref{sec:review} we review the free fermion, the KdV charges, and the known modular properties of GGE partition functions. In Section~\ref{sec:ggetwopt} we compute the exact two-point function of the fermion in a KdV GGE on the torus, for each spin structure and for several KdV charges. In Section~\ref{sec:section4} we determine its modular transformation order by order in the chemical potential, identify the dressing of the fermions, and rewrite it in terms of nested OPEs with the KdV currents. Motivated by this, we state the conjecture for a generic chiral algebra in Section~\ref{sec:6}. In Sections~\ref{proof} and ~\ref{sec:6.1}, we use Zhu's recursion relations to establish it at the lowest explicit power of the modular parameter, to all orders in the chemical potential, and to test the remaining terms. We end with a summary and outlook, and the appendices contain some additional technical details.

\section{Free fermions, KdV charges and generalised Gibbs ensembles}
\label{sec:review}

This section collects the background material that will be useful to set up the problem. We begin with the free Majorana fermion on the plane, the cylinder,  and the torus and then recall how its torus two-point function behaves under the modular group in the absence of any deformation. We then review the KdV charges of the free fermion and the recently obtained modular properties of generalised Gibbs ensembles of a chiral algebra. 

Throughout the paper, we use $q=e^{2\pi i\tau}$ with $\operatorname{Im}\tau>0$, and we work with the holomorphic (chiral) half of the theory only. For an operator $\mathcal O$ acting on the Hilbert space $\mathcal H_s$ of a sector $s$, we denote the unnormalised trace by
\begin{equation}
\label{eq:bracketdef}
\langle\mathcal O\rangle_{s,\epsilon}
=\Tr_{\mathcal H_s}\!\Big(\epsilon^{F}\,\mathcal O\;q^{L_0-\frac{c}{24}}\Big),
\qquad \epsilon=\pm1,
\end{equation}
where $\epsilon^{F}\equiv1$ for $\epsilon=+1$ and $\epsilon^{F}\equiv(-1)^F$ for $\epsilon=-1$, with $(-1)^F$ the fermion parity defined below. When the sector and the value of $\epsilon$ are not important, we simply write $\langle\mathcal O\rangle_\tau$ for $\Tr\big(\mathcal O\,q^{L_0-c/24}\big)$. We write normalised expectation values explicitly as ratios.

\subsection{The free Majorana fermion}
\label{sec:fermion}

The free Majorana fermion in two Euclidean dimensions has central charge $c=1/2$. It contains a holomorphic field $\psi(z)$ of conformal weight $1/2$, with operator product expansion
\begin{equation}
\label{psipsiope}
\psi(z_1)\psi(z_2)=\frac{1}{z_1-z_2}+\dots\,,
\end{equation}
and an antiholomorphic partner $\bar\psi(\bar z)$ with the same OPE. The two chiral halves decouple, and from now on we keep only~$\psi$.

Around the origin of the plane, the fermion can be either single-valued, $\psi(e^{2\pi i}z)=\psi(z)$, or antiperiodic, $\psi(e^{2\pi i}z)=-\psi(z)$. These two choices define the Neveu--Schwarz (NS) and Ramond (R) sectors, respectively, with mode expansions
\begin{equation}
\label{eq:fermionmode}
\begin{aligned}
\psi_{\mathrm R}(z)&=\sum_{m\in\mathbb Z}\psi_m\,z^{-m-1/2}\,,\\
\psi_{\mathrm{NS}}(z)&=\sum_{m\in\mathbb Z+1/2}\psi_m\,z^{-m-1/2}\,.
\end{aligned}
\end{equation}
The modes obey the anticommutation relations
\begin{equation}
\label{eq:psipsianti}
\{\psi_m,\psi_n\}=\delta_{m+n,0}\,.
\end{equation}
We also introduce the fermion parity $(-1)^F$, which squares to one and anticommutes with every mode, $\{\psi_m,(-1)^F\}=0$. A state is called bosonic if its eigenvalue under $(-1)^F$ is $+1$ and fermionic if it is $-1$.

In both sectors the ground state is annihilated by all positive modes, $\psi_m\ket{0}_s=0$ for $m>0$, and is bosonic, $(-1)^F\ket{0}_s=\ket{0}_s$. The NS sector has no zero mode and therefore a unique ground state, the $SL(2)$-invariant vacuum.

The Ramond sector requires separate care. Here~\eqref{eq:psipsianti} gives $\psi_0^2=\tfrac12$, so $\psi_0$ cannot annihilate the ground state. Instead, $\psi_0$ maps $\ket{0}_{\mathrm R}$ to a second ground state, which we normalise as
\begin{equation}
\ket{1}_{\mathrm R}=\sqrt2\,\psi_0\ket{0}_{\mathrm R}\,,
\qquad
(-1)^F\ket{1}_{\mathrm R}=-\ket{1}_{\mathrm R}\,.
\end{equation}
The two states are degenerate, since $[L_0,\psi_0]=0$. We work with the parity-graded Ramond Hilbert space generated from both of them.

In both sectors the Hilbert space $\mathcal H_s$, with $s\in\{\mathrm R,\mathrm{NS}\}$, is spanned by the states
\begin{equation}
\psi_{-m_k}\cdots\psi_{-m_2}\psi_{-m_1}\ket{a}_s\,,
\qquad m_k>\dots>m_2>m_1>0\,,\quad m_i\in\mathbb Z+\alpha_s\,,
\end{equation}
where $\alpha_{\mathrm{NS}}=\tfrac12$ and $\alpha_{\mathrm R}=0$, and where $a=0$ in the NS sector while $a\in\{0,1\}$ in the R sector. The inequalities are strict because $\psi_{-m}^2=0$ for $m\neq0$.

The pair $(s,\epsilon)$ specifies the spin structure on the torus: $s$ fixes the periodicity of the fermion around the spatial circle, and $\epsilon$ fixes that around the thermal circle. An ordinary trace, $\epsilon=+1$, makes the fermion antiperiodic around the thermal circle, whereas inserting $(-1)^F$, $\epsilon=-1$, makes it periodic. Of the four combinations, three give non-vanishing partition functions $\langle1\rangle_{s,\epsilon}$. In the Ramond sector with a parity insertion, the two ground states cancel in the trace,
\begin{equation}
\langle 1\rangle_{\mathrm R,-}
=\Tr_{\mathcal H_{\mathrm R}}\!\Big((-1)^Fq^{L_0-c/24}\Big)=0\,,
\end{equation}
so the normalised expectation values in the $(\mathrm R,-)$ sector are undefined. We therefore restrict to the sectors
\begin{equation}
(s,\epsilon)\in\{(\mathrm R,+),\,(\mathrm{NS},+),\,(\mathrm{NS},-)\}.
\end{equation}

\subsection{The two-point function on the torus and its modular properties}
\label{sec:section2}

We describe the torus by the cylinder coordinate $u$, with $z=e^u$ and identifications
\begin{equation}
u\sim u+2\pi i\sim u+2\pi i\tau\,.
\end{equation}
Since the fermion has conformal weight $\tfrac12$, its cylinder representative is
\begin{equation}
\psi(u)\equiv e^{u/2}\psi(e^u)=\sum_{r\in\mathbb Z+\alpha_s}\psi_r\,e^{-ru}\,.
\end{equation}
It is therefore periodic around the spatial circle in the Ramond sector and antiperiodic in the Neveu--Schwarz sector. We define the normalised two-point function in the sector $(s,\epsilon)$ as
\begin{equation}
\label{eq:twoptdefmuzero}
F_{s,\epsilon}(u;\tau)
=\frac{\langle\psi(u_1)\psi(u_2)\rangle_{s,\epsilon}}{\langle1\rangle_{s,\epsilon}}\,,
\qquad u=u_2-u_1\,.
\end{equation}
Translation invariance along the spatial circle ensures that it depends on the insertion points only through $u$. The only non-vanishing contractions of the modes are $\langle\psi_r\psi_{-r}\rangle$, which are fixed by the anticommutation relation~\eqref{eq:psipsianti}, the commutator $[L_0,\psi_r]=-r\psi_r$ and the cyclicity of the trace. The result is
\begin{equation}
\label{eq:twoptmuzero}
F_{s,\epsilon}(u;\tau)
=\sum_{r\in\mathbb Z+\alpha_s}\frac{e^{ru}}{1+\epsilon\,q^{r}}\,,
\qquad (s,\epsilon)\in\{(\mathrm R,+),(\mathrm{NS},+),(\mathrm{NS},-)\}\,.
\end{equation}
The series converges absolutely in the strip
\begin{equation}
-2\pi\operatorname{Im}\tau<\operatorname{Re}u<0\,,
\end{equation}
and defines the correlator elsewhere by analytic continuation.

The mode sums can be resummed in terms of Jacobi theta functions \cite{DiFrancesco:1997nk},
\begin{equation}
\label{eq:fermion-theta}
F_{s,\epsilon}(u;\tau)
=-\frac{1}{2\pi i}\,
\frac{\vartheta_{\nu(s,\epsilon)}\!\left(\frac{u}{2\pi i}\middle|\tau\right)\vartheta_1'(0|\tau)}
{\vartheta_{\nu(s,\epsilon)}(0|\tau)\;\vartheta_1\!\left(\frac{u}{2\pi i}\middle|\tau\right)}\,,
\qquad
\nu(\mathrm R,+)=2,\quad \nu(\mathrm{NS},+)=3,\quad \nu(\mathrm{NS},-)=4\,.
\end{equation}
Here the theta functions have periods $1$ and $\tau$ in their first argument, and the prime denotes differentiation with respect to it. The prefactor and the argument follow from the cylinder convention $z=e^u$ and $u=u_2-u_1$. As $u\to0$ the correlator has the expected singularity, $F_{s,\epsilon}(u;\tau)\sim-1/u$.

\subsubsection{Modular transformations}
Under the modular $S$ transformation,
\begin{equation}
S:\qquad \tau\to-\frac1\tau\,,\qquad u\to\frac u\tau\,,
\end{equation}
the two cycles of the torus are exchanged, and the transformation properties of the theta functions give
\begin{equation}
\begin{aligned}
F_{\mathrm R,+}\!\Big(\frac u\tau;-\frac1\tau\Big)&=\tau\,F_{\mathrm{NS},-}(u;\tau)\,,\\
F_{\mathrm{NS},+}\!\Big(\frac u\tau;-\frac1\tau\Big)&=\tau\,F_{\mathrm{NS},+}(u;\tau)\,,\\
F_{\mathrm{NS},-}\!\Big(\frac u\tau;-\frac1\tau\Big)&=\tau\,F_{\mathrm R,+}(u;\tau)\,.
\end{aligned}
\end{equation}
Thus the $(\mathrm{NS},+)$ correlator is mapped to itself, while $(\mathrm R,+)$ and $(\mathrm{NS},-)$ are exchanged. This is the expected action of $S$ on the spin structures. Collecting the three correlators in the vector
\begin{equation}
\mathbf F(u;\tau)=\begin{pmatrix}F_{\mathrm{NS},+}(u;\tau)\\F_{\mathrm{NS},-}(u;\tau)\\F_{\mathrm R,+}(u;\tau)\end{pmatrix},
\end{equation}
The relations above take the form
\begin{equation}
\label{eq:fermion-modular-S}
\mathbf F\!\Big(\frac u\tau;-\frac1\tau\Big)
=\tau\begin{pmatrix}1&0&0\\0&0&1\\0&1&0\end{pmatrix}\mathbf F(u;\tau)\,.
\end{equation}
Similarly, under the $T$ transformation $\tau\to\tau+1$, with $u$ fixed, the factor $q^r$ in~\eqref{eq:twoptmuzero} acquires the phase $e^{2\pi ir}$. This phase is $-1$ for $r\in\mathbb Z+\tfrac12$ and $+1$ for $r\in\mathbb Z$, so $T$ exchanges $(\mathrm{NS},+)$ and $(\mathrm{NS},-)$ and leaves $(\mathrm R,+)$ invariant:
\begin{equation}
\label{eq:fermion-modular-T}
\mathbf F(u;\tau+1)=\begin{pmatrix}0&1&0\\1&0&0\\0&0&1\end{pmatrix}\mathbf F(u;\tau)\,.
\end{equation}
The phases picked up by the theta functions in these transformations cancel between the numerator and the denominator of the normalised correlator. The matrices above therefore describe the transformation of the normalised two-point functions, not of the chiral characters. In summary, the three correlators form a vector-valued meromorphic Jacobi form of weight one (and index zero, since the Gaussian factors in the theta-function transformations also cancel). This is the reference against which the KdV-deformed correlators of the following sections will be compared.

\subsection{KdV charges of the free fermion}
\label{sec:kdv}

The free fermion has an infinite set of mutually commuting local conserved charges, the KdV charges $I_{2k-1}$ with $k=1,2,\dots$, which are the zero modes of currents $J_{2k}$ of spin $2k$ \cite{Downing:2021mfw}. On the cylinder, the spatial circle is the A-cycle $u\sim u+2\pi i$, and
\begin{equation}
\label{eq:KdVCdef}
I_{2k-1}=\int_0^{2\pi i}\frac{du}{2\pi i}\,J_{2k}(u)\,.
\end{equation}
The first current is the stress tensor, $J_2=T=\tfrac12(\partial\psi\,\psi)$. Mapping it from the plane to the cylinder gives
\begin{equation}
I_1=\oint\Big(z^2T(z)-\frac{c}{24}\Big)\frac{dz}{2\pi i\,z}=L_0-\frac c{24}\,,
\end{equation}
so the thermal weight itself is the Gibbs weight of the first KdV charge, $q^{L_0-c/24}=e^{2\pi i\tau I_1}$. A GGE adds the higher charges to this weight.

For the free fermion, the higher currents are, up to total derivatives, bilinears in $\psi$,
\begin{equation}
J_{2k}(u)=\tfrac12\,\big(\partial^{2k-1}\psi\,\psi\big)(u)\quad\text{mod }\partial\,.
\end{equation}
Total derivatives do not contribute to the charges~\eqref{eq:KdVCdef}.\footnote{The OPEs below depend on the choice of representative. They hold for the symmetric representative $J_{2k}(u)=2^{-2k}(\partial_{u_1}-\partial_{u_2})^{2k-1}\,{:}\psi(u_1)\psi(u_2){:}\big|_{u_1=u_2=u}$, built from the cylinder fermion. For $k=1$, this is the stress tensor. Adding a total derivative to $J_{2k}$ changes the poles of its OPEs but not the charges.} In terms of the fermion modes, the charges are
\begin{equation}
I_{2k-1}=\sum_{m>0}m^{2k-1}\,\psi_{-m}\psi_m-c^{\mathrm R/\mathrm{NS}}_{2k-1}\,,
\end{equation}
where the constants can be obtained by zeta-function regularisation \cite{Downing:2021mfw},
\begin{equation}
c^{\mathrm R}_{2k-1}=\tfrac12\,\zeta(1-2k)\,,\qquad
c^{\mathrm{NS}}_{2k-1}=\tfrac12\Big(\big(\tfrac12\big)^{2k-1}-1\Big)\zeta(1-2k)\,.
\end{equation}
(For $k=1$ these reproduce $I_1=L_0-\tfrac1{48}$ in the NS sector.) The constants multiply the identity and drop out of all normalised correlators. What matters below is the commutator of a charge with a fermion mode. Since $2k-1$ is odd, it takes the same form for positive and negative modes:
\begin{equation}
\label{eq:psiIcom}
[I_{2k-1},\psi_m]=-m^{2k-1}\psi_m\,,\qquad m\in\mathbb Z+\alpha_s\,.
\end{equation}
Consequently, the KdV charges commute with each other and with $L_0$ and $(-1)^F$.

The modular transformation of correlators in the GGE is governed by the singular terms in the OPEs between the currents and the other fields in the correlator. For the free fermion, we need two sets of them. The OPEs between KdV currents, in the cylinder coordinate $u$, read
\begin{equation}
\label{eq:kdvcurope}
J_{2k}(u_1)J_{2m}(u_2)\sim\sum_{n}\frac{(J_{2k}J_{2m})_n(u_2)}{(u_1-u_2)^n}\,,
\end{equation}
with the first two poles
\begin{equation}
\label{eq:kdvcurope1}
\begin{aligned}
(J_{2k}J_{2m})_1&=(2k-1)\,\partial J_{2k+2m-2}+\text{higher total derivatives}\,,\\
(J_{2k}J_{2m})_2&=(2k+2m-2)\,J_{2k+2m-2}+\text{total derivatives}\,.
\end{aligned}
\end{equation}
The OPE of a KdV current with the fermion has, for its first two poles,
\begin{equation}
\label{eq:Jpsiope}
J_{2k}(u_1)\psi(u_2)=\dots+\frac{2k-1}{2}\,\frac{\partial^{2k-2}\psi(u_2)}{(u_1-u_2)^2}+\frac{\partial^{2k-1}\psi(u_2)}{u_1-u_2}+\dots\,,
\end{equation}
where $\partial=\partial_{u_2}$ and the dots stand for the higher-order poles and the regular terms. The same expressions hold on the plane, with $u$ replaced by $z$.

\subsection{Modular properties of generalised Gibbs ensembles}
\label{sec:ggemod}

We now recall what is known about the modular properties of GGEs \cite{Downing:2025huv, Ashok:2026cyi}. Let $W(u)$ be a holomorphic current of conformal weight $w$, and let
\begin{equation}
W_0=\int_0^{2\pi i}\frac{du}{2\pi i}\,W(u)
\end{equation}
be its zero mode, i.e.\ its integral over the A-cycle of the torus. Turning on a chemical potential $\mu$ for this charge defines the generalised partition function
\begin{equation}
\label{eq:ggecha}
Z(\tau,\mu)=\big\langle e^{\mu W_0}\big\rangle_\tau\,,
\end{equation}
where the trace is taken over a fixed representation of the underlying symmetry algebra. This function is no longer modular covariant in the usual sense. The $S$ transformation now acts on the chemical potential as well as on the modulus,
\begin{equation}
S:\qquad \tau\to-\frac1\tau\,,\qquad \mu\to\frac{\mu}{\tau^{w}}\,,
\end{equation}
and the transformed partition function is again a GGE partition function, but for a different charge. It was proposed in \cite{Downing:2025huv}, for the spin-$3$ current of the $\mathcal W_3$ algebra, and derived in \cite{Ashok:2026cyi} for a general chiral current, that
\begin{equation}
\label{OGeqn}
S\Big[\big\langle e^{\mu W_0}\big\rangle_\tau\Big]
:=\Big\langle e^{\frac{\mu}{\tau^{w}}W_0}\Big\rangle_{-1/\tau}
=\big\langle e^{\mu\,\mathcal W_0}\big\rangle_\tau\,,
\end{equation}
where the equality holds at the level of the asymptotic expansion in $\mu$. Here $\mathcal W_0$ is the zero mode of a field $\mathcal W$ defined by the series
\begin{equation}
\label{calWdefn}
\mathcal W=\sum_{n=0}^{\infty}\frac{x^n}{2^n\,n!}\,[W^{n+1}],
\qquad x=\frac{\mu}{2\pi i\tau}\,.
\end{equation}
The composite fields $[W^{n+1}]$ have weight $(n+1)w-2n$ and are generated recursively by
\begin{equation}
\label{Wnrecursionv2}
[W^{n+1}]=(WW)_2\,\frac{\partial}{\partial W}\,[W^{n}]\,,\qquad [W^1]=W\,.
\end{equation}
Here $(AB)_2$ denotes the coefficient of the second-order pole in the OPE of $A$ with $B$, and $\partial/\partial W$ is a replacement operator. It acts as a derivation on the composite field: it replaces each factor of $W$ in turn by $(WW)_2$ and sums over the replacements. For example,
\begin{equation}
[W^2]=(WW)_2\,,\qquad [W^3]=\big((WW)_2W\big)_2+\big(W(WW)_2\big)_2\,.
\end{equation}
The field $\mathcal W$ depends on $\tau$ and $\mu$ only through $x$.
The extension to several chiral deformations is discussed in Section 7 of \cite{Ashok:2026cyi}. 

\section{The free fermion two-point function in a GGE}
\label{sec:ggetwopt}

We now compute the normalised fermion two-point function in the GGE obtained by adding a single KdV charge to the thermal weight. For an integer $k>1$ and each of the three sectors $(s,\epsilon)$, we define
\begin{equation}
\label{eq:twoptdef}
F_{s,\epsilon}(u,\tau,\mu)
=\frac{\big\langle\psi(u_1)\psi(u_2)\,e^{\mu I_{2k-1}}\big\rangle_{s,\epsilon}}
{\big\langle e^{\mu I_{2k-1}}\big\rangle_{s,\epsilon}}\,,\qquad u=u_2-u_1\,,
\end{equation}
where $\mu$ is the chemical potential conjugate to $I_{2k-1}$. We exclude $k=1$ because $I_1$ is already part of the thermal weight. At $\mu=0$, this reduces to the undeformed correlator~\eqref{eq:twoptdefmuzero}. As before, the correlator depends on $u_1$ and $u_2$ only through $u=u_2-u_1$. At finite chemical potential, this definition applies whenever the traces converge, and the denominator is nonzero.

We denote the full density operator by
\begin{equation}
\rho=\epsilon^{F}\,e^{\mu I_{2k-1}}\,q^{L_0-\frac{c}{24}}\,.
\end{equation}
Conjugation with $\rho$ rescales the modes. This follows from $[L_0,\psi_r]=-r\psi_r$, from the commutator~\eqref{eq:psiIcom}, and from $(-1)^F\psi_r(-1)^F=-\psi_r$:
\begin{equation}
\rho\,\psi_r\,\rho^{-1}=\epsilon\,q^{-r}\,e^{-\mu r^{2k-1}}\,\psi_r\,.
\end{equation}
Since $\rho$ commutes with $L_0$, the trace $\Tr(\rho\,\psi_r\psi_{s})$ vanishes unless $r+s=0$. For $s=-r$, we use the anti-commutation relation~\eqref{eq:psipsianti} and the cyclicity of the trace to write
\begin{equation}
\Tr(\rho\,\psi_r\psi_{-r})
=\Tr(\rho)-\Tr(\psi_r\,\rho\,\psi_{-r})
=\Tr(\rho)-\epsilon\,q^{r}e^{\mu r^{2k-1}}\Tr(\rho\,\psi_r\psi_{-r})\,,
\end{equation}
and hence
\begin{equation}
\frac{\Tr(\rho\,\psi_r\psi_{-r})}{\Tr(\rho)}=\frac{1}{1+\epsilon\,q^{r}e^{\mu r^{2k-1}}}\,.
\end{equation}
This also covers the Ramond zero mode, $r=0$, where it gives $\psi_0^2=\tfrac12$ for $\epsilon=+1$. Summing over the modes with the weights $e^{-ru_1+ru_2}$ from $\psi(u)=\sum_r\psi_re^{-ru}$ gives
\begin{equation}
\label{eq:twoptkdv}
F_{s,\epsilon}(u,\tau,\mu)=\sum_{r\in\mathbb Z+\alpha_s}\frac{e^{ru}}{1+\epsilon\,q^{r}\,e^{\mu r^{2k-1}}}\,,
\end{equation}
with $\alpha_{\mathrm R}=0$ and $\alpha_{\mathrm{NS}}=\tfrac12$. This reduces to~\eqref{eq:twoptmuzero} at $\mu=0$. For all three sectors, the deformation enters only through the replacement $q^r\to q^r e^{\mu r^{2k-1}}$ in the denominator.

\subsection{Extension to several charges}

The same argument applies to a GGE with several KdV charges. We add to the weight a chemical potential $\mu_{2j+1}$ for each charge $I_{2j+1}$ with $j\ge1$, i.e.\ the charges $I_3,I_5,\dots$ (the charge $I_1$ is already accounted for by $q$). The normalised two-point function is
\begin{equation}
\label{eq:twoptmkdvdef}
F_{s,\epsilon}(u,\tau,\{\mu_i\})
=\frac{\big\langle\psi(u_1)\psi(u_2)\,e^{\sum_{j\ge1}\mu_{2j+1}I_{2j+1}}\big\rangle_{s,\epsilon}}
{\big\langle e^{\sum_{j\ge1}\mu_{2j+1}I_{2j+1}}\big\rangle_{s,\epsilon}}\,.
\end{equation}
The replacement $\mu\,r^{2k-1}\to\sum_{j\ge1}\mu_{2j+1}r^{2j+1}$ in the derivation above gives
\begin{equation}
\label{eq:twoptmkdv}
F_{s,\epsilon}(u,\tau,\{\mu_i\})
=\sum_{r\in\mathbb Z+\alpha_s}\frac{e^{ru}}{1+\epsilon\,q^{r}\,e^{\sum_{j\ge1}\mu_{2j+1}r^{2j+1}}}\,.
\end{equation}
As for the partition function, the dependence of these correlators on the chemical potentials is understood as an asymptotic expansion around $\mu=0$. At $\mu\neq0$, the correlators are no longer Jacobi forms. The remainder of the paper addresses how they transform under $ S$.

\section{The modular transformation of the two-point function}
\label{sec:section4}
In this section, we study the fermion two-point function deformed
by a single KdV charge. We begin by doing an asymptotic expansion of \eqref{eq:twoptkdv}  in the chemical potential around $\mu=0$: 
\begin{equation}
\begin{aligned}
    F_{\alpha,\epsilon}(u,\tau,\mu)&\sim\sum_{n\ge0}\sum_{m\in\mathbb{Z}+\alpha}\frac{\mu^n}{n!}(\partial_u)^{n(2k-2)}(q\partial_q)^n\left(\frac{e^{mu}}{1+\epsilon q^m}\right)\\
&=\sum_{n\ge0}\frac{\mu^n}{n!}(\partial_u)^{n(2k-2)}(q\partial_q)^n F_{\alpha,\epsilon}(u,\tau,0)~,
\end{aligned} 
\end{equation}
where $F_{\alpha,\epsilon}(u,\tau,0)$ is the undeformed two-point function. As we know from our discussion in section \ref{sec:section2}, the undeformed two-point function behaves as a weight one index zero vector-valued Jacobi form: 
\begin{equation}
    F_s(\frac{u}{\tau},-\frac{1}{\tau},0)=\tau S_{sr}F_r(u,\tau,0),~~~s\in \{(\NS,+),(\NS,-),(\R,+)\},
\end{equation}
where the matrix $S$ is given in \eqref{eq:fermion-modular-S} and the repeated index is summed over. The vanishing index means that their modular transformation
contains no exponential factor depending on the insertion
coordinate. Thus, the modular $S$-transformation of the deformed two-point
function reduces to determining how the $\tau$ and $u$-derivatives
of the undeformed weight-one Jacobi form transform. We perform this in the following way: let $f(u,\tau)$ be a Jacobi form of weight $w$ and index zero. Under the modular $S$ transformation, we get $S(u,\tau)\rightarrow (u^{\prime},\tau^{\prime})=\Big(\frac{u}{\tau},-\frac{1}{\tau}\Big)$ and the Jacobi form transforms as $f(u^{\prime},\tau^{\prime})=\tau^w f(u,\tau)$. We want to find the modular transform of 
\begin{equation}
    S[\partial^{m}_\tau\partial^{n}_uf(u,\tau)]\rightarrow \partial^{m}_{\tau^{\prime}}\partial^{n}_{u^{\prime}}f(u^{\prime},\tau^{\prime})=f^{(n,m)}(u^{\prime},\tau^{\prime})~,
\end{equation}
which is given by
\begin{equation}
\label{eq:MTJacobideri}
     f^{(n,m)}(u^{\prime},\tau^{\prime})=  \sum_{\substack{r\ge 0,\, l\ge n \\ l+r\le n+m}}\frac{m!n!}{l!r!}\binom{l}{n} \binom{-1 + w + m + n}{n+m-l-r}u^{l-n}\tau^{w + m + n + r}f^{(l,r)}(u,\tau)
\end{equation}
The details of the derivation of the above equation are presented in Appendix \ref{app:A}. Now, using \eqref{eq:MTJacobideri}, we can write
\begin{equation}
   \begin{aligned}
   \label{eq:MTtwoptff}
        F_s(\frac{u}{\tau},-\frac{1}{\tau},\frac{\mu}{\tau^{2k}})&\sim\sum_{n\geq0}\frac{\mu^n}{(2\pi i)^n}\sum_{l=0}^{n}\sum_{r=0}^{n-l} \frac{1}{l!r!}\binom{n(2k-1)}{n-l-r} u^{l}\tau^{-n+1+r}\partial_{\tau}^r\partial_u^{l+n(2k-2)}S_{st}F_{t}(u,\tau,0),
   \end{aligned}
\end{equation}
where we have put $w=1$. The matrix $S$ is the same as in \eqref{eq:fermion-modular-S}. We can write the first few terms of the above expansion as
\begin{equation}
     F_s(\frac{u}{\tau},-\frac{1}{\tau},\frac{\mu}{\tau^{2k}})\sim\tau S_{rs}F_{r}(u,\tau,0)+\frac{\mu}{2\pi i}\Big(\tau\partial_{\tau}\partial_u^{2k-2}+u\partial^{2k-1}_u+(2k-1)\partial^{2k-2}_u\Big) S_{st}F_{t}(u,\tau,0)+\ldots
\end{equation}
Since the $S$ matrix is independent of $u$ and $\tau$, it commutes with the differential operators. We can then write \eqref{eq:MTtwoptff} as 
\begin{equation}
\label{eq:MTtwoptffn}
     F_s(\frac{u}{\tau},-\frac{1}{\tau},\frac{\mu}{\tau^{2k}})\sim \tau S_{rs}\mathcal{M}F_{r}(u,\tau,0),
\end{equation}
where $\mathcal{M}$ is the differential operator acting on the undeformed two-point function. Thus, perturbatively, the modular $S$-transform changes the spin structure of the deformed two-point function. They do not give a sum over multiple spin structures due to the special structure of the $S$ matrix in the free fermion theory. This is an important observation which we will exploit below.

\subsection{The effective current and the dressing operator}
\label{sec:fermion-effective-ensemble}

Our eventual goal is to obtain the modular transformation of GGE correlators in a generic CFT. Thus, although \eqref{eq:MTtwoptff} gives the free-fermion result explicitly, its present form does not suggest a direct generalisation. Following \cite{Downing:2025huv, Ashok:2026cyi}, we therefore seek to recast it in terms of nested OPEs involving the KdV current and the external fields, providing a formulation that can be more easily generalised to other rational models. We propose the following ansatz for the modular-transformed free fermion two-point function:
\begin{equation}
\label{eq:ffansatz}
    F^{ans}_{s}(\frac{u}{\tau},-\frac{1}{\tau},\frac{\mu}{\tau^{2k}})\sim~~\tau\frac{S_{sr}\Tr_{r}\Big([\mathcal{U}(x)\psi(u_1)\psi(u_2)]q^{L_0-\frac{c}{24}}~e^{\mu  { {\cal W}_0 }}\Big)}{S_{sr}\Tr_{r}\Big(q^{L_0-\frac{c}{24}}~e^{\mu  { {\cal W}_0 }}\Big)},
\end{equation}
where $x=\frac{\mu}{2\pi i\tau}$ and the matrix $S$ denotes the modular transformation matrix for the unnormalised traces. In the free-fermion spin-structure basis, each row has only one nonzero entry. Consequently, the sums in the numerator and denominator select the same sector, and the common matrix element cancels. The resulting transformation of the normalised correlator therefore agrees \eqref{eq:fermion-modular-S}, as discussed below \eqref{eq:MTtwoptffn}. We call $\mathcal U(x)$ the dressing operator and
$\mathcal W(x)$ the effective current. The latter is defined
by \eqref{calWdefn}, with $W$ replaced by the KdV current
$J_{2k}$. Its zero mode, $\mathcal W(x)_0$, will be called
the effective charge.

In \eqref{eq:ffansatz}, every quantity is fixed apart from $\mathcal{U}(x)$. We will demand that the ansatz in \eqref{eq:ffansatz} matches the explicit result in \eqref{eq:MTtwoptff} to compute $\mathcal{U}(x)$ order by order in chemical potential $\mu$. Here, $\mathcal{U}(x)$ acts on the product $\psi(u_1)\psi(u_2)$, with contributions from its action at both insertion points, and reduces to the identity at $\mu=0$.  Based on this, we propose the following expansion:  

\begin{equation}
    \mathcal{U}(x)=1+\sum_{n\ge 1} x^n U_{n}. 
\end{equation}
We would like to write $\mathcal{U}(x)$ as a local action on the external fermions generated by nested OPEs with the KdV currents. Since these currents are bilinear, their singular OPEs with a fermion contain only its coordinate descendants. Their action therefore admits a representation involving derivatives with respect to the insertion coordinates. On a translation-invariant two-point function, these reduce to derivatives with respect to $u=u_2-u_1$.

We therefore distinguish the field action  $\mathcal{U}(x)$ from its coordinate differential representation $\widehat{\mathcal{U}}(x)$, and write
\begin{equation}
\label{eq:hatu}
   \widehat{\mathcal{U}}(x)=1+\sum_{n\ge 1} x^n \widehat{U}_n. 
\end{equation}
We will compute $\widehat{U}_n$ using the free fermion result in \eqref{eq:MTtwoptff}. Since each row of the free-fermion modular matrix selects a single spin structure, we may work sector by sector and suppress the sector label. We therefore write \eqref{eq:ffansatz} as    

\begin{equation}
\label{eq:ffansatz2}
    F^{ans}(\frac{u}{\tau},-\frac{1}{\tau},\frac{\mu}{\tau^{2k}})\sim~~\tau\frac{\Tr\Big(\mathcal{U}(x)\psi(u_1)\psi(u_2)q^{L_0-\frac{c}{24}}~e^{\mu  { {\cal W}_0 }}\Big)}{\Tr\Big(q^{L_0-\frac{c}{24}}~e^{\mu  { {\cal W}_0 }}\Big)}.
\end{equation}
Using the coordinate differential representation, we can write the above equation as 
\begin{equation}
\label{eq:ffansatz3}
    F^{ans}(\frac{u}{\tau},-\frac{1}{\tau},\frac{\mu}{\tau^{2k}})\sim~~\tau~\widehat{\mathcal{U}}(x)\frac{\Tr\Big(\psi(u_1)\psi(u_2)q^{L_0-\frac{c}{24}}~e^{\mu  { {\cal W}_0 }}\Big)}{\Tr\Big(q^{L_0-\frac{c}{24}}~e^{\mu  { {\cal W}_0 }}\Big)}.
\end{equation}
Next, we evaluate $\mathcal{W}$ using its defining relations \eqref{calWdefn} and \eqref{Wnrecursionv2}, together with the OPEs between the KdV currents given in \eqref{eq:kdvcurope} and \eqref{eq:kdvcurope1}. We get
\begin{equation}
\label{eq:calwff}
 \mathcal{W}(x)=\sum_{n=0}^{\infty}\frac{1}{2^nn!}x^nc(n,k)J_{2k+n(2k-2)}+\partial \mathcal{O},
\end{equation}
where 
\begin{equation}
    c(n,k)=\frac{2^n n!}{n+1}\binom{(n+1)(2k-1)}{n}.
\end{equation}
  Using the form of $\mathcal{W}$ for the KdV current $J_{2k}$, we can write the normalised two-point function which appears in \eqref{eq:ffansatz3} as 
\begin{equation}
\label{eq:fftwoptcalwdef}
  \frac{\Tr\Big(\psi(u_1)\psi(u_2)q^{L_0-\frac{c}{24}}~e^{\mu  { {\cal W}_0 }}\Big)}{\Tr\Big(q^{L_0-\frac{c}{24}}~e^{\mu  { {\cal W}_0 }}\Big)}=\sum_{m\in \mathbb{Z}+\alpha}  \frac{e^{mu}}{1+\epsilon \exp[\mu\sum_{j\ge0}a_j x^j m^{d+1+jd}]q^m}~,
\end{equation}
where $d=2k-2$. Here $a_j$ is given by  
\begin{equation}
    a_j=\frac{1}{j+1}\binom{(j+1)(2k-1)}{j}.
\end{equation}
Therefore, an asymptotic expansion in $\mu$ in the right-hand side of \eqref{eq:fftwoptcalwdef} gives us

   \begin{equation}
\label{eq:fftwoptcalwdef1}
  \frac{\Tr\Big(\psi(u_1)\psi(u_2)q^{L_0-\frac{c}{24}}~e^{\mu  { {\cal W}_0 }}\Big)}{\Tr\Big(q^{L_0-\frac{c}{24}}~e^{\mu  { {\cal W}_0 }}\Big)}\sim \sum_{r\ge0}\frac{\tau^r}{r!}\Big(\sum_{j\ge1}x^jA_j\Big)^r F_{\alpha,\epsilon}(u,\hat{\tau},0)|_{\hat{\tau}=\tau}
\end{equation} 
where 
\begin{equation}
    A_j=a_{j-1}\partial_{\hat{\tau}}\partial^{jd}_u,~~~\forall ~~j\ge 1.
\end{equation}
We have introduced the auxiliary variable $\hat{\tau}$ to stress the fact that the derivatives in $A_j$ do not act on $x$. Now, we use the following partial Bell polynomial identity 
\begin{equation}
    \frac{1}{r!}\Big(\sum_{j\ge1}x^jA_j\Big)^r =\sum_{n\ge r}\frac{x^n}{n!}B_{n,r}(A_1,2!A_2,\ldots, (n-r+1)!A_{n-r+1})
\end{equation}
and write \eqref{eq:fftwoptcalwdef1} as 
\begin{equation}
  \frac{\Tr\Big(\psi(u_1)\psi(u_2)q^{L_0-\frac{c}{24}}~e^{\mu  { {\cal W}_0 }}\Big)}{\Tr\Big(q^{L_0-\frac{c}{24}}~e^{\mu  { {\cal W}_0 }}\Big)}\sim \sum_{n\ge 0}\frac{x^n}{n!}\sum_{r\ge0}^{n}\tau^r~B_{n,r}(A_1,2!A_2,\ldots, (n-r+1)!A_{n-r+1})F_{\alpha,\epsilon}(u,\hat{\tau},0)|_{\hat{\tau}=\tau}
\end{equation}

Next, we use the series representation of the operator $\widehat{\mathcal{U}}(x)$ defined in \eqref{eq:hatu} and we can write the ansatz for the modular $S$-transformed fermion two point function in \eqref{eq:ffansatz3} as 
\begin{equation}
\label{eq:ffansatz4}
    \begin{aligned}
        F^{ans}(\frac{u}{\tau},&-\frac{1}{\tau},\frac{\mu}{\tau^{2k}})\sim~~\tau~\sum_{n\ge 0}\sum_{p=0}^{n}\frac{x^n\widehat{U}_{p}}{(n-p)!}\\\times&\sum_{r\ge0}^{n-p}\tau^r~B_{n-p,r}(A_1,2!A_2,\ldots, (n-p-r+1)!A_{n-p-r+1})F_{\alpha,\epsilon}(u,\hat{\tau},0)|_{\hat{\tau}=\tau}
    \end{aligned}
\end{equation}
Now, we demand that the ansatz for the modular $ S$- transformed fermion two-point function in \eqref{eq:ffansatz3} is the same as the explicit result in \eqref{eq:MTtwoptff}. Comparing the coefficient of $x^n$, we have 
\begin{equation}
    \begin{aligned}
        &\sum_{p=0}^{n}\frac{\widehat{U}_{p}}{(n-p)!}\sum_{r\ge0}^{n-p}\tau^r~B_{n-p,r}(A_1,2!A_2,\ldots, (n-p-r+1)!A_{n-p-r+1})F_{\alpha,\epsilon}(u,\hat{\tau},0)|_{\hat{\tau}=\tau}\\=&\sum_{l=0}^{n}\sum_{r=0}^{n-l} \frac{1}{l!r!}\binom{n(2k-1)}{n-l-r} u^{l}\tau^{r}\partial_{\tau}^r\partial_u^{l+n(2k-2)}F_{\alpha,\epsilon}(u,\tau,0)
    \end{aligned}
\end{equation}
Setting $r=0$ and using the following identity satisfied by the partial Bell polynomial: 
\begin{equation}
  B_{n-p,0}(A_1,2!A_2,\dots,(n-p+1)!A_{(n-p+1)})=\delta_{n-p,0},
\end{equation}
we get
\begin{equation}
\label{eq:hatUff}
  \widehat{U}_{n}=\sum_{l=0}^{n}  \frac{1}{l!}\binom{n(2k-1)}{n-l} u^{l}\partial_u^{l+n(2k-2)}.
\end{equation}
Explicitly, the first few operators take the form:
\begin{equation}
    \begin{aligned}
      \widehat{U}_0=&1~,\\ 
           \widehat{U}_1=&(2k-1)\partial_u^{(2k-2)}+u\partial_u^{(2k-2)+1}~,\\
           \widehat{U}_{2}=&(-1 + 2 k) (-3 + 4 k)\partial_u^{2(2k-2)}+2(2k-1)u\partial_u^{2(2k-2)+1}+\frac{1}{2}u^2\partial_u^{2(2k-2)+2}~.
    \end{aligned}
\end{equation} 

\subsection{Verification to all orders}
\label{sec:fermion-all-order-matching}
Now that we have found $\widehat{\mathcal{U}}(x)$, we proceed to show that the ansatz for the free fermion two-point function in \eqref{eq:ffansatz} reproduces the explicit result for the $S$-transform of the free fermion two-point function given in \eqref{eq:MTtwoptff}. We define the scalar coefficients
\begin{equation}
   b_{N,r}:=\frac{1}{N!}B_{N,r}(a_0,2!a_1,\ldots),~~~~~~b_{0,0}=1. 
\end{equation}
Since $A_j=a_{j-1}\partial_{\hat{\tau}}\partial^{jd}_{u}$, using the homogeneous property of Bell polynomials, we can therefore write 
\begin{equation}
\label{eq:bellhomo}
    \frac{1}{N!}B_{N,r}(A_1,2!A_2,\ldots)=b_{N,r}\partial^r_{\hat{\tau}}\partial^{Nd}_{u}
\end{equation}
Using the above equation and the form of $\widehat{U}(x)$ given in \eqref{eq:hatUff}, we can write \eqref{eq:ffansatz4} as 
\begin{equation}
\label{eq:ffansatz5}
    \begin{aligned}
        F^{ans}(\frac{u}{\tau},&-\frac{1}{\tau},\frac{\mu}{\tau^{2k}})\sim~~\tau~\sum_{n\ge 0}x^n\sum_{p=0}^{n}\sum_{l=0}^{p}  \frac{1}{l!}\binom{p(2k-1)}{p-l} u^{l}\sum_{r\ge0}^{n-p}\tau^r~b_{n-p,r}\partial^r_{\hat{\tau}}\partial^{(nd+l)}_{u}F_{\alpha,\epsilon}(u,\hat{\tau},0)|_{\hat{\tau}=\tau}
    \end{aligned}
\end{equation}
Upon some reordering of the sums, the above equation can be  written as 
\begin{equation}
\label{eq:ffansatz6}
    \begin{aligned}
        F^{ans}(\frac{u}{\tau},&-\frac{1}{\tau},\frac{\mu}{\tau^{2k}})\sim~~\tau~\sum_{n\ge 0}x^n\sum_{l=0}^{n}\sum_{r\ge0}^{n-l} \sum_{p\ge l}^{n-r} \frac{1}{l!}\binom{p(2k-1)}{p-l} u^{l}\tau^r~b_{n-p,r}\partial^r_{\hat{\tau}}\partial^{(nd+l)}_{u}F_{\alpha,\epsilon}(u,\hat{\tau},0)|_{\hat{\tau}=\tau}
    \end{aligned}
\end{equation}
We next use the following identity (proved in Appendix \ref{app:Appendix2}):
\begin{equation}
    \sum_{p=l}^{n-r}\binom{p(2k-1)}{p-l}b_{n-p,r}=\frac{1}{r!} \binom{n(2k-1)}{n-l-r},~~~~l+r\le n~,
\end{equation}
and write \eqref{eq:ffansatz6} as 
\begin{equation}
\label{eq:ffansatz7}
    \begin{aligned}
        F^{ans}(\frac{u}{\tau},&-\frac{1}{\tau},\frac{\mu}{\tau^{2k}})\sim~~\tau~\sum_{n\ge 0}x^n\sum_{l=0}^{n}\sum_{r\ge0}^{n-l} \frac{1}{r!l!} \binom{n(2k-1)}{n-l-r}u^{l}\tau^r~\partial^r_{\tau}\partial^{(nd+l)}_{u}F_{\alpha,\epsilon}(u,\tau,0),
    \end{aligned}
\end{equation}
which matches with \eqref{eq:MTtwoptff}. We have proved that the ansatz in \eqref{eq:ffansatz} involving the dressing operator reproduces the explicit result for the $S$-transform of the free fermion two-point function, provided the collection of dressing operators  $\left\{U_n\right\}$ is given by  \eqref{eq:hatUff}.

\subsection{The dressing operator in terms of OPEs}
\label{sec:fermion-ope-dressing}
The free fermion result has allowed us to identify a differential operator representation for the dressing operator. But, in order to lift this result to a generic CFT, we would like to express the operator so obtained in terms of nested OPEs. For a current $X$ and a local field $\phi$,
define
\begin{equation}
\label{eq:external-ope-action}
    \mathcal{D}_X\phi(u_a)
    :=
    u_a(X\phi)_1(u_a)+(X\phi)_2(u_a).
\end{equation}
Here $(X\phi)_j$ denotes the coefficient of $(w-u_a)^{-j}$ in
the OPE $X(w)\phi(u_a)$. On the two external fermions, the action is
\begin{equation}
\label{eq:external-ope-action-two-fields}
\begin{aligned}
    \mathcal{D}_X[\psi(u_1)\psi(u_2)]
    ={}&
    (\mathcal{D}_X\psi(u_1))\psi(u_2)
    +\psi(u_1)(\mathcal{D}_X\psi(u_2)).
\end{aligned}
\end{equation}
Since the KdV currents are bosonic, no additional fermionic sign
appears. Using the OPE between KdV currents and fermion fields given in \eqref{eq:Jpsiope}, we can therefore write 

\begin{equation}
\label{eq:kdv-action-one-fermion}
    \mathcal{D}_{J_{2a}}\psi(u_i)
    =
    \left[
        u_i\partial_{u_i}^{2a-1}
        +\frac{2a-1}{2}\partial_{u_i}^{2a-2}
    \right]\psi(u_i).
\end{equation}
For a translation-invariant two-point function, with
$u=u_2-u_1$, we have
$\partial_{u_1}=-\partial_u$ and
$\partial_{u_2}=\partial_u$. The coordinate differential representation of the combined action is therefore
\begin{equation}
\label{eq:kdv-action-coordinate}
    L_{J_{2a}}
    =
    u\partial_u^{2a-1}
    +(2a-1)\partial_u^{2a-2}.
\end{equation}
Here $L_{J_{2a}}$ is the differential operator representation of the action of $ \mathcal{D}_{J_{2a}}$ on two external fermion fields. $L_{J_{2a}}$ acts on the correlator whereas $ \mathcal{D}_{J_{2a}}$ acts on the fields. Using \eqref{calWdefn}, we can write the differential operator representation of the action of $ \mathcal{D}_{\mathcal{W}(x)}$ as

\begin{equation}
\label{eq:calW-action-coordinate}
    L_{\mathcal{W}(x)}
    =
    \sum_{j\geq0}a_jx^j
    \left[
        u\partial_u^{(j+1)d+1}
        +\bigl((j+1)d+1\bigr)
        \partial_u^{(j+1)d}
    \right].
\end{equation}
Here, we have ignored the total derivative that appears in \eqref{eq:calwff}. In Appendix \ref{app:total-derivatives}, we show that the total derivatives are always at least second order in nature and therefore do not affect our results. The differential operator $ L_{\mathcal{W}(x)}$ acts on the two-point function. To understand its action, it is enough to study its action on the function $e^{mu}$, where $u=u_2-u_1$ and $m$ is an arbitrary integer. The action of $ L_{\mathcal{W}(x)}$ is given by
\begin{equation}
\label{eq:calLW-on-exponential}
\begin{aligned}
    L_{\mathcal{W}(x)}e^{mu}
    &=
    \sum_{j\geq0}a_jx^j
    \left[
        um^{(j+1)d+1}
        +\bigl((j+1)d+1\bigr)m^{(j+1)d}
    \right]e^{mu}
\end{aligned}
\end{equation}
Similarly, using the form of $\widehat{\mathcal{U}}(x)$ given in \eqref{eq:hatUff}, we can find the action of the operator $\widehat{\mathcal{U}}(x)$ on $e^{mu}$ gives
\begin{equation}
\label{eq:Uhat-on-exponential}
    \widehat{\mathcal{U}}(x)e^{mu}
    =
    \sum_{n\geq0}x^n
    \sum_{\ell=0}^{n}
    \frac{1}{\ell!}
    \binom{n(d+1)}{n-\ell}
    u^\ell m^{nd+\ell}e^{mu},
\end{equation}
Using the Leibniz rule, we can write
\begin{equation}
\label{eq:Uhat-derivative-series}
    \widehat{\mathcal{U}}(x)e^{mu}
    =
    \sum_{n\geq0}
    \frac{x^n}{n!}
    \partial_m^n
    \left[m^{n(d+1)}e^{mu}\right].
\end{equation}

Now that we have understood the action $\widehat{\mathcal{U}}(x)$ and $ L_{\mathcal{W}(x)}$ on $e^{mu}$, we now want to understand how these two operations are related. For this reason, we introduce the unique formal power series $t=t(m,x)$, which satisfies
\begin{equation}
\label{eq:auxiliary-t-definition}
    t=m+xt^{d+1},
    \qquad
    t(m,0)=m.
\end{equation}
Defining the following notations
\begin{equation}
     t_m:=\left.\frac{\partial t}{\partial m}\right|_x,
    \qquad
    t_x:=\left.\frac{\partial t}{\partial x}\right|_m,
\end{equation}
allows us to write
\begin{equation}
\label{eq:auxiliary-t-derivatives}
    t_m=\frac{1}{1-(d+1)xt^d},
    \qquad
    t_x=\frac{t^{d+1}}{1-(d+1)xt^d}
    =t^{d+1}t_m.
\end{equation}
Now let us consider a function $H(t)$ and applying Lagrange inversion formula \cite{Gessel_2016} on $H(t)$, we get
\begin{equation}
\label{eq:Haux}
    H(t)
    =
    H(m)
    +
    \sum_{n\geq1}
    \frac{x^n}{n!}
    \partial_m^{n-1}
    \left[m^{n(d+1)}H'(m)\right].
\end{equation}
Differentiating both sides with respect to $m$, we obtain
\begin{equation}
    \begin{aligned}
        t_m H^{\prime}(t)=& H^{\prime}(m)
    +
    \sum_{n\geq1}
    \frac{x^n}{n!}
    \partial_m^{n}
    \left[m^{n(d+1)}H'(m)\right].
    \end{aligned}
\end{equation}
Now we choose some special form of the auxiliary field $H(t)$. If $H(t)$ is a function which satisfies $H^{\prime}(t)=e^{ut}$, then we have 
\begin{equation}
t_m e^{ut}=\sum_{n\geq0}
    \frac{x^n}{n!}
    \partial_m^{n}
    \left[m^{n(d+1)}e^{mu}\right]= \widehat{\mathcal{U}}(x)e^{mu}
\end{equation}
On the other hand, if we choose $H(z)=z$ and $H^{\prime}(z)=1$, then from \eqref{eq:Haux}, we have
\begin{equation}
\label{eq:texp}
  \begin{aligned}
       t =m+\sum_{n\geq1}\frac{x^n}{n!} \partial_m^{n-1}\left[m^{n(d+1)}\right]
    =m+\sum_{n\geq1}
    \frac{x^n}{n}
    \binom{n(d+1)}{n-1}
    m^{nd+1}.
  \end{aligned}
\end{equation}
and using $t^{d+1}=\frac{t-m}{x}$, we get
\begin{equation}
\label{eq:tdexp}
   t^{d+1}
    =
    \sum_{j\geq0}
    a_jx^jm^{(j+1)d+1}.
\end{equation}
Now, we can write the action of $ L_{\mathcal{W}(x)}$ on $e^{mu}$ given in \eqref{eq:calLW-on-exponential} as
\begin{equation}
\label{eq:calW-on-exponential}
\begin{aligned}
    L_{\mathcal{W}(x)}e^{mu}
    &=
    \sum_{j\geq0}a_jx^j
    \left[
        um^{(j+1)d+1}
        +\bigl((j+1)d+1\bigr)m^{(j+1)d}
    \right]e^{mu}
    \\
    &=
    \left[ut^{d+1}+\partial_m(t^{d+1})\right]e^{mu}=
    \partial_m\left(t^{d+1}e^{mu}\right).
\end{aligned}
\end{equation}
Since $\widehat{\mathcal{U}}$ differentiates only $u$, it commutes
with $\partial_m$ and multiplication by $t(m,x)^{d+1}$. Hence
\begin{equation}
\label{eq:Uhat-evolution-proof}
\begin{aligned}
    \widehat{\mathcal{U}}(x)L_{\mathcal{W}(x)}e^{mu}
    &=
    \partial_m
    \left[t^{d+1}\widehat{\mathcal{U}}(x)e^{mu}\right]
    =
    \partial_m\left[t^{d+1}t_m e^{ut}\right]
    \\
    &=
    \partial_m\left[t_xe^{ut}\right]
   =
    \partial_x\left[t_me^{ut}\right]
    =
    \widehat{\mathcal{U}}'e^{mu}.
\end{aligned}
\end{equation}
Here we have used the fact that $t_x=t^{d+1}t_m$, which we showed in \eqref{eq:auxiliary-t-derivatives}. This analysis therefore proves
\begin{equation}
\label{eq:Uhat-evolution}
   \frac{d\widehat{\mathcal{U}}(x)}{dx}
 =
    \widehat{\mathcal{U}}(x)L_{\mathcal{W}(x)}.
\end{equation}
We now translate this coordinate differential identity into an
identity for field actions. Consider the following OPE between two fields $X$ and $\phi$:
\begin{equation}
    X(w)\phi(u)\sim \frac{(X\phi)_1(u)}{w-u}+ \frac{(X\phi)_2(u)}{
    (w-u)^2}+\dots
\end{equation}
Differentiating the OPE with respect to
the insertion coordinate gives
\begin{equation}
    (X\partial\phi)_1=\partial(X\phi)_1,
    \qquad
    (X\partial\phi)_2=(X\phi)_1+\partial(X\phi)_2.
\end{equation}
Then we can write using the definition \eqref{eq:external-ope-action}
\begin{equation}
\label{eq:D-action-descendants}
\begin{aligned}
    \mathcal{D}_X(\partial\phi)
    &=
    u\partial(X\phi)_1+(X\phi)_1+\partial(X\phi)_2
    \\
    &=
    \partial\left[u(X\phi)_1+(X\phi)_2\right]
    =
    \partial(\mathcal{D}_X\phi).
\end{aligned}
\end{equation}
Thus, if $\mathcal{D}_Y\phi=L_Y\phi$, then
\begin{equation}
\label{eq:field-coordinate-order-reversal}
    \mathcal{D}_X\mathcal{D}_Y\phi
    =
    \mathcal{D}_X(L_Y\phi)
    =
    L_Y(\mathcal{D}_X\phi)
    =
    L_YL_X\phi.
\end{equation}
In particular, $\mathcal{D}_{\mathcal{W}(x)}~\mathcal{U}(x)$ is represented
by $\widehat{\mathcal{U}}(x)L_{\mathcal{W}(x)}$. Consequently, \eqref{eq:Uhat-evolution} gives, in its action on the
external fermion insertions,
\begin{equation}
\label{eq:U-field-evolution}  
    \frac{d\mathcal{U}(x)}{dx}
    = \mathcal{D}_{\mathcal{W}(x)}\mathcal{U}(x),
    \qquad
    \mathcal{U}(0)=1.
\end{equation}
Substituting
\begin{equation}
    \mathcal{U}(x)=\sum_{n\geq0}x^nU_n,
    \qquad
    \mathcal{D}_{\mathcal{W}(x)}
    =
    \sum_{j\geq0}
    \frac{x^j}{2^jj!}\mathcal{D}_{[W^{j+1}]},
\end{equation}
and comparing coefficients of $x^n$, we obtain
\begin{equation}
\label{eq:U-field-recursion}
    (n+1)U_{n+1}
    =
    \sum_{j=0}^{n}
    \frac{1}{2^jj!}
    \mathcal{D}_{[W^{j+1}]}U_{n-j},
    \qquad U_0=1.
\end{equation}
The first coefficients are
\begin{equation}
\label{eq:U-field-first-coefficients}
\begin{aligned}
    U_1={}&\mathcal{D}_W,\\
    U_2={}&
    \frac12\mathcal{D}_W^2
    +\frac14\mathcal{D}_{[W^2]},\\
    U_3={}&
    \frac16\mathcal{D}_W^3
    +\frac1{12}\mathcal{D}_W\mathcal{D}_{[W^2]}
    +\frac16\mathcal{D}_{[W^2]}\mathcal{D}_W
    +\frac1{24}\mathcal{D}_{[W^3]}.
\end{aligned}
\end{equation}

The operator extracted from the exact free-fermion modular expansion
therefore admits the nested-OPE representation
\eqref{eq:U-field-recursion}, provided the local-current identification
used in \eqref{eq:calW-action-coordinate} holds. Appendix \ref{app:total-derivatives} establishes the required local-current identification up to second total derivatives, which leave both the zero mode and the external-field action unchanged, thereby completing the free-fermion verification. This formulation
motivates the corresponding conjecture for a generic CFT.

\section{A conjecture for rational chiral CFTs}
\label{sec:6}

The free-fermion calculation suggests that the modular
transformation of GGE correlators can be expressed through
an effective current $\mathcal{W}(x)$ and a transformation of the external
fields with an operator $\mathcal{U}(x)$. We now formulate this proposal for rational chiral CFTs.

We use the cylinder coordinate $z=e^u$, with
$u\sim u+2\pi i$, and consider a bosonic holomorphic current
$W$ of integer spin $w$. Its zero mode is
\begin{equation}
\label{eq:W-zero-mode}
    W_0=\int_0^{2\pi i}\frac{du}{2\pi i}\,W(u).
\end{equation}
The current and its local representative are fixed
throughout. All statements below are understood as formal
power series in the chemical potential. In Section \ref{sec:section4}, we considered normalised two-point functions, written explicitly as ratios of unnormalised traces. We now formulate the conjecture for unnormalised correlators. As in \eqref{eq:bracketdef}, we denote by $\langle\cdots\rangle_{\tau,s}$the unnormalised torus trace over the Hilbert space $\mathcal H_s$, where $\tau$ is the complex structure of the torus and (s) labels the sector. We obtain normalised correlators by dividing by the corresponding GGE partition function.

Consider the correlation function of $N$ chiral primary fields in the GGE. For this purpose, let us define 
\begin{equation}
    \Phi=\prod_{a=1}^{N}\phi_{a}(u_a)
\end{equation}
where $h_a$ is the conformal dimension of the field $\phi_a$. 
We then define
\begin{equation}
\label{eq:Phi-S-definition}
    \hat{\Phi}=\prod_{a=1}^{N}\phi_a(u_a/\tau).
\end{equation}
Under the modular $ S$- transform, we have
\begin{equation}
    \tau\rightarrow -\frac{1}{\tau},\qquad u\rightarrow\frac{u}{\tau},\qquad \mu\rightarrow\frac{\mu}{\tau^w}. 
\end{equation}
The torus correlator transforms as
\begin{equation}
S\left[\left\langle
 \Phi\,
e^{\mu W_0}
\right\rangle_{\tau,s}\right]:= \left\langle
 \hat{\Phi}\,
e^{\mu \frac{W_0}{\tau^{w}}}
\right\rangle_{-\frac{1}{\tau},s}~,
\end{equation}
and we propose that 
\begin{equation}
\left\langle
 \hat{\Phi}\,
e^{\mu \frac{W_0}{\tau^{w}}}
\right\rangle_{-\frac{1}{\tau},s}
\sim
\tau^{h_\Phi}
S_{sr}\left\langle
e^{\mu\mathcal{W}(x)_0}\,
\mathcal{U}(x)\,
\Phi
\right\rangle_{\tau,r},
\qquad
x=\frac{\mu}{2\pi i\tau}.
\label{eq:fullconj}
\end{equation}
Here, the operator $\mathcal{W}(x)$ is the same effective current that appeared in the modular $S$-transform of the generalised character \cite{Ashok:2026cyi} (see equation \eqref{calWdefn} and \eqref{Wnrecursionv2}), and the dressing operator ${\cal U}(x)$ is the one defined recursively by the relation \eqref{eq:U-field-evolution}.

In the equation, $h_\Phi=\sum_{a=1}^{N}h_a$, and $S_{sr}$ is a particular representation of the modular $S$ matrix. As before, the formula for the $S$-transform is valid only at the level of the asymptotic expansion of the chemical potential. 

To compare with the normalised two-point function in \eqref{eq:ffansatz}, we divide \eqref{eq:fullconj} by the corresponding modular transformation of the GGE partition function. This gives
\begin{equation}
\frac{
\left\langle e^{\mu W_0/\tau^w}\hat{\Phi}\right\rangle_{-1/\tau,s}
}{
\left\langle e^{\mu W_0/\tau^w}\right\rangle_{-1/\tau,s}
}
\sim
\tau^{h_\Phi}
\frac{
 S_{sr}
\left\langle e^{\mu\mathcal W(x)_0}
\bigl[\mathcal U(x)\Phi\bigr]\right\rangle_{\tau,r}
}{
 S_{sr}
\left\langle e^{\mu\mathcal W(x)_0}\right\rangle_{\tau,r}
}.
\end{equation}
Thus, the normalised transformation involves sums over sectors in both the numerator and denominator, whereas \eqref{eq:fullconj} contains a single sum because it describes an unnormalised correlator. For the free-fermion two-point function, $h_\Phi=1$, and this expression reduces to the form of \eqref{eq:ffansatz}.

\subsection{Towards the proof of the conjecture}
\label{proof}
We start by expanding both sides of the conjecture as a power series in $x$. We can write the left-hand side of the conjecture \eqref{eq:fullconj} as 
\begin{equation}
    \left\langle
 \hat{\Phi}\,
e^{\mu \frac{W_0}{\tau^{w}}}
\right\rangle_{-\frac{1}{\tau},s}=  \sum_{n=0}^{\infty} (2\pi i\, \tau^{1-w})^n \frac{x^n}{n!} \langle W_0^n \hat{\Phi} \rangle_{-\frac{1}{\tau},s}.
\end{equation}
We now express the modular transformation in terms of torus correlators containing zero modes and local fields. From the definition of zero modes \eqref{eq:W-zero-mode}, we can write
\begin{equation}
   \langle W_0^n \Phi \rangle_{\tau,s} =\frac{1}{(2\pi i)^n}\int_{0}^{2\pi i}\prod_{i=1}^{n}dw_i\langle W(w_1)\dots W(w_n) \Phi \rangle_{\tau,s}.
\end{equation}
Under modular transformation, it has been proven by Zhu in \cite{zhu:1996} (see Theorem 5.3.2) that such a torus correlator transforms homogeneously as long as the state corresponding to $W$ transforms homogeneously under the scaling transformations on the plane.   
So, for the  local field
\footnote{Although Theorem 5.3.2 in \cite{zhu:1996} is proven for primaries, the postscript of the theorem clearly states the stronger result quoted here.} $W(w)$ and $\phi_a(u)$ of definite weight $w$ and $h_a$, under the S-transformation, the correlator picks up a factor $\tau^{nw+h_{\phi}}$. Using this theorem, we have 
\begin{equation}
  \begin{aligned}
      S:  \langle W(w_1)\dots W(w_n) \Phi \rangle_{\tau,s}&\rightarrow \langle W(w^{\prime}_1)\dots W(w^{\prime}_n) \hat{\Phi}\rangle _{-\frac{1}{\tau},s}\\
      &=\tau^{nw+h_\Phi}S_{sr}\langle W(w_1)\dots W(w_n) \Phi\rangle _{\tau,r},
  \end{aligned}
\end{equation}
where $w^{\prime}_i=\frac{w_i}{\tau}$. So, for the integrated correlator, we obtain 
\begin{equation}
  \begin{aligned}
       \langle W_0^n \hat{\Phi} \rangle_{-\frac{1}{\tau},s} &=\frac{1}{(2\pi i)^n}\int_{0}^{2\pi i}\prod_{i=1}^{n}dw^{\prime}_i\langle W(w^{\prime}_1)\dots W(w^{\prime}_n) \hat{\Phi} \rangle_{-\frac{1}{\tau},s}\\
       &=\frac{1}{(2\pi i)^n}\tau^{nw+h_\Phi-n}\int_{0}^{2\pi i\tau}\prod_{i=1}^{n}dw_iS_{sr}\langle W(w_1)\dots W(w_n) \Phi \rangle_{\tau,r}
  \end{aligned}
\end{equation}
Thus, the left-hand side of the conjecture can be written in terms of the integrated correlators as 
\begin{equation}
\label{eq:explhs}
    \left\langle
 \hat{\Phi}\,
e^{\mu \frac{W_0}{\tau^{w}}}
\right\rangle_{-\frac{1}{\tau},s}=  \sum_{n=0}^{\infty} \tau^{h_\Phi} \frac{x^n}{n!}\int_{0}^{2\pi i\tau}\prod_{i=1}^{n}dw_iS_{sr}\langle W(w_1)\dots W(w_n) \Phi \rangle_{\tau,r},
\end{equation}
where we are integrating over the $B$ cycle of the torus.
Now, if we expand the right-hand side of \eqref{eq:fullconj}, we get

\begin{equation}
\label{eq:cojexprhs}
\allowdisplaybreaks
    \begin{aligned}
&\tau^{h_\Phi}\sum_r S_{sr}
\left\langle
    e^{\mu\mathcal W(x)_0}[\mathcal U(x)\Phi]
\right\rangle_{\tau,r}=\tau^{h_\Phi}\sum_r S_{sr}
\Biggl\{
    \langle\Phi\rangle_{\tau,r}
+x\Bigl[
    (2\pi i\tau)\langle W_0\Phi\rangle_{\tau,r}
    +\langle U_1\Phi\rangle_{\tau,r}
\Bigr]
\\
&\quad+x^2\Biggl[
    \frac{(2\pi i\tau)^2}{2!}
        \langle W_0^2\Phi\rangle_{\tau,r}
    +(2\pi i\tau)\left(
        \langle W_0[U_1\Phi]\rangle_{\tau,r}
        +\frac12\langle[W^2]_0\Phi\rangle_{\tau,r}
    \right)
    +\langle U_2\Phi\rangle_{\tau,r}
\Biggr]
\\
&\quad+x^3\Biggl[
    \frac{(2\pi i\tau)^3}{3!}
        \langle W_0^3\Phi\rangle_{\tau,r}
    +\frac{(2\pi i\tau)^2}{2!}
    \left(
        \langle W_0^2[U_1\Phi]\rangle_{\tau,r}
        +\frac12
        \left\langle
            \bigl(W_0[W^2]_0+[W^2]_0W_0\bigr)\Phi
        \right\rangle_{\tau,r}
    \right)
\\
&\qquad
    +(2\pi i\tau)\left(
        \langle W_0[U_2\Phi]\rangle_{\tau,r}
        +\frac12\langle[W^2]_0[U_1\Phi]\rangle_{\tau,r}
        +\frac18\langle[W^3]_0\Phi\rangle_{\tau,r}
    \right)
    +\langle U_3\Phi\rangle_{\tau,r}
\Biggr]
+\dots
\Biggr\}.
\end{aligned}
\end{equation}
Now, if we isolate the coefficient of $x^n$ from the above equation, we have
\begin{equation}
    \begin{aligned}
    \label{eq:exprhsxn}
    \left[\tau^{h_\Phi}
S_{sr}\left\langle
e^{\mu\mathcal{W}(x)_0}\,
\mathcal{U}(x)\,
\Phi
\right\rangle_{\tau,r}\right]|_{x^n}&= \tau^{h_\Phi}S_{sr}\Big[ \frac{(2\pi i \tau)^n}{n!}\langle W_0^n\Phi \rangle_{\tau,r} +\dots + \langle U_n\Phi \rangle_{\tau,r} 
        \Big].
    \end{aligned}
\end{equation}
Upon comparing the $x^n$ coefficient in \eqref{eq:explhs} and \eqref{eq:exprhsxn}, we have 
\begin{equation}
\label{eq:conjequiv}
   \frac{1}{n!} \int_{0}^{2\pi i\tau}\prod_{i=1}^{n}dw_iS_{sr}\langle W(w_1)\dots W(w_n) \Phi \rangle_{\tau,r}=S_{sr}\Big[\frac{(2\pi i \tau)^n}{n!}\langle W_0^n\Phi \rangle_{\tau,r} +\dots + \langle U_n\Phi \rangle_{\tau,r} 
        \Big]
\end{equation}
This gives an equivalent formulation of the conjecture and
provides a concrete target for the proof. As a consistency
check, we verify this identity up to $n=3$ in
Appendix~\ref{app:D} using the Zhu recursion relation. However, evaluating the integrated
correlator at arbitrary order $n$, while retaining all powers
of $\tau$, is a challenging task. Thus we take a different route.

We first observe in \eqref{eq:conjequiv} that the RHS at order $\tau^0$ is just the trace of the action of $U_n$ on the external insertions $\Phi$. If we define the order $\tau^0$ part of the integrated correlator as 
\begin{equation}
    \mathcal{G}^n:=\frac{1}{n!}\Bigg[\int_{0}^{2\pi i\tau}\prod_{i=1}^{n}dw_iS_{sr}\langle W(w_1)\dots W(w_n) \Phi \rangle_{\tau,r}\Bigg]\Bigg|_{\tau^0},
\end{equation}
then we can write 
\begin{equation}
\label{eq:conjequiv2}
    \mathcal{G}^n=S_{sr}\langle U_n\Phi \rangle_{\tau,r},
\end{equation}
 where $U_n$ should satisfy the recursion relation \eqref{eq:U-field-recursion}. Here, $\big|_{\tau^0}$ denotes the coefficient that does not have
explicit power of $\tau$ outside the torus traces, with their
full $q=e^{2\pi i\tau}$ dependence retained. It does not denote
evaluation at $\tau=0$.

Let us now explain what \eqref{eq:conjequiv2} establishes for all $n$ and what it does not. Let us assume that the modular transformed correlator satisfies the conjecture for some fields  $\mathcal{U}(x)$ and $\mathcal{W}(x)$, where $\mathcal{U}(x)=1+\sum_{n\ge1}U_nx^n$ and the coefficients do not have explicit $\tau$ dependence. \eqref{eq:conjequiv2} tells us that the term that does not have any explicit $\tau$ dependence does not depend on $\mathcal{W}(x)$ and only depends on $U_n$. Thus \eqref{eq:conjequiv2} completely determines the operator $U_n$ which satisfies the recursion relation \eqref{eq:U-field-recursion}. Therefore, we can construct the complete dressing function by only studying the order $\tau^0$ part of the integrated correlator. The terms with explicit powers of $\tau$ in the integrated correlator \eqref{eq:conjequiv} have zero-mode insertions, and they are not constrained by \eqref{eq:conjequiv2}. We have checked them up to $n=3$ in the appendix \ref{app:D}. We therefore prove the conjecture at order $\tau^0$ for all orders in $\mu$. Terms that involve higher powers of $\tau$ depend on the structure of $\mathcal{W}(x)$ found for the partition function and are not proved here. 

However, the form of $\mathcal{W}(x)$ is well motivated. In the conjecture, if we choose $\Phi=\mathds{1}$, the left-hand side reduces to the generalised character in the GGE. The right-hand side should reproduce the 
 modular-transformed generalised character in the GGE as given by \cite{Downing:2025huv, Ashok:2026cyi}. The action of $\mathcal{U}(x)$ on $\Phi$ when $\Phi=\mathds{1}$ becomes $\mathcal{U}(x)\Phi=1$. Thus, under the assumption that $\mathcal{W}(x)$ does not depend on the choice of insertion $\Phi$, it should have the same form as in \cite{Downing:2025huv, Ashok:2026cyi}. That being said, we do not claim to have a complete proof of the conjecture, and a complete all-order proof remains an open question which is outside the scope of this work.

\subsection{Deriving the dressing operator}
\label{sec:6.1}

In this section, we prove the following statement:
\begin{equation}
\label{eq:conjequiv3}
    \mathcal{G}^n=S_{sr}\langle U_n\Phi \rangle_{\tau,r},
\end{equation}
where $U_n$ should satisfy the recursion relation \eqref{eq:U-field-recursion}. Since the following identities are established sector
by sector and are linear in the torus traces, we suppress
the sector labels and restore the modular sum at the end.

We first denote the integrated correlator as 
\begin{equation}
\label{eq:intco1}
   I_n[\Phi]:=\frac{1}{n!}\int_{0}^{2\pi i\tau}\prod_{i=1}^{n}dw_i\langle W(w_1)\dots W(w_n) \Phi \rangle_{\tau}   
\end{equation}
To avoid coincident singularities, we integrate
the currents along separated representatives of the torus
$B$-cycle. We choose the following small real displacements satisfying
\begin{equation}
\label{eq:contour-displacements}
    0=\epsilon_1<\epsilon_2<\cdots<\epsilon_n<2\pi,
\end{equation}
and define the oriented contours
\begin{equation}
\label{eq:shifted-B-contours}
    \mathcal C_j:
    \qquad
    w_j(t)=i\epsilon_j+2\pi i\tau t,
    \qquad 0\leq t\leq1.
\end{equation}
Thus $\mathcal C_1$ runs from $0$ to $2\pi i\tau$,
while the remaining contours are displaced in the
positive $A$-cycle direction. Then we can write the integrated correlator as 
\begin{equation}
\label{eq:ordered-integrated-correlator}
    I_n[\Phi]:=
    \frac{1}{n!}\int_{\mathcal C_n}dw_n\cdots
    \int_{\mathcal C_1}dw_1\,
    \left\langle
        W(w_1)\cdots W(w_n)\Phi
    \right\rangle_\tau,
\end{equation}
where the integrations are performed starting with
$w_1$ and ending with $w_n$.

The external insertion points and their periodic images
are chosen away from the contours. The displacements
are sufficiently small that no external insertion lies
in the strips between adjacent contours. This contour
ordering is retained throughout the calculation;
deformations across current or external-field insertions
must include the corresponding OPE residues.

We will begin with \eqref{eq:intco1} and use the Zhu recursion to get
\begin{equation}
    \begin{aligned}
      I_n[\Phi]=&\frac{1}{n!}(2\pi i\tau)\int_0^{2\pi i\tau}d^{n-1} w\langle W_0W(w_2)\dots W(w_n)\Phi\rangle\\
        &+\frac{1}{n!}\int_0^{2\pi i\tau}d^{n} w~\sum_{i=1}^{N}\sum_{m\ge0}\mathcal{P}_{m+1}(e^{(u_i-w_1)})\langle W(w_2)\dots W(w_n)(W[m]\Phi)_i\rangle\\
        &+\frac{1}{n!}\int_0^{2\pi i\tau}d^{n} w\sum_{j\ge 2}\sum_{m\ge0}\mathcal{P}_{m+1}(e^{(w_j-w_1)})\langle W(w_2)\dots (W[m]W(w_j))\dots W(w_n)\Phi\rangle
    \end{aligned}
\end{equation}
Here, $(W[m]\Phi)_i$ denotes the action of $W[m]$ on the
$i$-th insertion only:
\begin{equation}
    (W[m]\Phi)_i
:=\phi_1(u_1)\cdots(W[m]\phi_i)(u_i)\cdots\phi_N(u_N).
\end{equation}
Next, we perform the $w_1$ integration. To do that, we use the integral identities given in \eqref{eq:revint}. Upon performing the integration, we use the following identity \footnote{We have proved this in Appendix \ref{app:Appendix3}.}

 \begin{equation}
 \label{eq:id}
     \sum_{j\ge2}^{n}\langle X_2(u_2)\ldots X_1[0]X_j(u_j)\ldots X_n(u_n)\rangle=0
 \end{equation}
 where $X_i(u_i)$ are local fields. Using this identity, we get
\begin{equation}
    \begin{aligned}
      I_n[\Phi]=&\frac{1}{n!}(2\pi i\tau)\int_0^{2\pi i\tau}d^{n-1} w\langle W_0W(w_2)\dots W(w_n)\Phi\rangle\\
        &+\frac{1}{n!}\int_0^{2\pi i\tau}d^{n-1} w\langle W(w_2)\dots W(w_n)\mathcal{D}_W\Phi\rangle\\
        &+\frac{1}{n!}\int_0^{2\pi i\tau}d^{n-1} w\sum_{j\ge 2}\langle W(w_2)\dots \Big(w_jW[0]W+W[1]W\Big)(w_j)\dots W(w_n)\Phi\rangle
    \end{aligned}
\end{equation}
We first observe that the first term in the above equation has a multiplicative $\tau$. Since we are only interested in the $\tau^0$ term, we can ignore that term. We can write
\begin{equation}
    \begin{aligned}
      &I_n[\Phi]=\frac{1}{n!}\int_0^{2\pi i\tau}d^{n-1} w\langle W(w_2)\dots W(w_n)\mathcal{D}_W\Phi\rangle\\
        &+\frac{1}{n!}\int_0^{2\pi i\tau}d^{n-1} w\sum_{j\ge 2}\langle W(w_2)\dots \Big(w_jW[0]W+W[1]W\Big)(w_j)\dots W(w_n)\Phi\rangle+\mathcal{O}(\tau)+\dots
    \end{aligned}
\end{equation}
We use the relation between the square bracket mode operation on a field and OPE, which is given by\cite{Ashok:2026cyi}\footnote{The relative factor of $(2\pi i)^n$ compared with
\cite{Ashok:2026cyi} arises from our different conventions
for the plane-to-cylinder map.}

\begin{equation}
   \langle \ldots A[n-1]B \ldots\rangle=\langle\ldots(AB)_n\dots\rangle.
\end{equation}
Upon the identification, we can therefore write
\begin{equation}
    \langle \ldots W[0]W \ldots\rangle=  \left\langle \ldots\Big(\frac{1}{2}\partial W[1]W+\partial^2\mathcal{O}\Big)\dots\right\rangle.
\end{equation}
Thus we have
\begin{equation}
\begin{aligned}
    &\frac{1}{n!}\int_0^{2\pi i\tau}d^{n-1} w\sum_{j\ge 2}\langle W(w_2)\dots w_j(W[0]W)(w_j)\dots W(w_n)\Phi\rangle  \\
    =&\frac{1}{2}\frac{1}{n!}\int_0^{2\pi i\tau}d^{n-1} w\sum_{j\ge 2}\langle W(w_2)\dots w_j\partial
    (W[1]W)(w_j)\dots W(w_n)\Phi\rangle +\text{higher derivatives }
\end{aligned}  
\end{equation}
We perform integration by parts in $w_j$ along its original
contour $\mathcal C_j$, keeping all other insertion
coordinates fixed. Writing the integrand as
$w_j\partial_{w_j}F_j(w_j)$, we obtain
\begin{equation}
\label{eq:integration-by-parts-fixed-contour}
    \int_{\mathcal C_j}dw_j\,w_j\partial_{w_j}F_j(w_j)
    =
    \left[w_jF_j(w_j)\right]_{i\epsilon_j}
        ^{i\epsilon_j+2\pi i\tau}
    -
    \int_{\mathcal C_j}dw_j\,F_j(w_j).
\end{equation}
The separated contours ensure that $F_j$ has no
singularities on the integration path. Since we do not
move $\mathcal C_j$ across another insertion, no contour
residues are generated. If $F_j$ is periodic along the
$B$-cycle, the boundary contribution simplifies to
\begin{equation}
\label{eq:integration-by-parts-boundary}
    \left[w_jF_j(w_j)\right]_{i\epsilon_j}
        ^{i\epsilon_j+2\pi i\tau}
    =
    2\pi i\tau\,F_j(i\epsilon_j).
\end{equation}
Thus periodicity does not remove the boundary term,
because the prefactor $w_j$ is not periodic.
Using the above discussion and the periodicity of the torus correlator, we can write 
\begin{equation}
   \begin{aligned}
       &\frac{1}{n!}\int_0^{2\pi i\tau}d^{n-1} w\sum_{j\ge 2}\langle W(w_2)\dots w_j(W[0]W)(w_j)\dots W(w_n)\Phi\rangle  \\
   =&-\frac{1}{n!}\frac{1}{2}\int_0^{2\pi i\tau}d^{n-1} w\sum_{j\ge 2}\langle W(w_2)\dots
    (W[1]W)(w_j)\dots W(w_n)\Phi\rangle +2\pi i\tau B
   \end{aligned}
\end{equation}
Here, $B$ denotes the contribution from the boundary term. Since $B$ comes with a $\tau$ prefactor, this term will not contribute to $\mathcal{G}_n$. Thus we have 
\begin{equation}
    \begin{aligned}
    \label{eq:Inrec1}
         I_n[\Phi]=&\frac{1}{n!}\int_0^{2\pi i\tau}d^{n-1} w\langle W(w_2)\dots W(w_n)\mathcal{D}_W\Phi\rangle\\
        +&\frac{1}{n!}\frac{1}{2}\int_0^{2\pi i\tau}d^{n-1} w\sum_{j\ge 2}\langle W(w_2)\dots (W[1]W)(w_j)\dots W(w_n)\Phi\rangle+\mathcal{O}(\tau)+\dots
    \end{aligned}
\end{equation}
We now define the following variational derivative operator
\begin{equation}
    \delta_W:=(WW)_2\partial_W,
\end{equation}
which acts on each occurrence of the local field W and replaces it with $(WW)_2$. A more formal definition is available in section 6.1 of \cite{Ashok:2026cyi}. Using this variational derivative, we can write \eqref{eq:Inrec1} as
\begin{equation}
\label{eq:Inrec2}
    nI_n[\Phi]=I_{n-1}[\Dc_W\Phi]+\frac{1}{2}\delta_W I_{n-1}[\Phi]+\mathcal{O}(\tau)
\end{equation}
with the initial condition 
\begin{equation}
    I_0[\Phi]=\langle \Phi\rangle\qquad \text{and}\qquad \delta_WI_0[\Phi]=0.
\end{equation}
Now using \eqref{eq:Inrec2}, we compute the $I_{[\Phi]}$ for first few values of $n$. We have
\begin{equation}
    \begin{aligned}
        I_1[\Phi]=&\langle \Dc_W\Phi\rangle+\mathcal{O}(\tau)\\
        I_2[\Phi]=&\frac{1}{4}\langle \Dc_{(WW)_2}\Phi\rangle+\frac{1}{2}\langle(\Dc_W)^2\Phi\rangle+\mathcal{O}(\tau)\\
         I_3[\Phi]=&\frac{1}{24}\langle \Dc_{[W^3]}\Phi\rangle+\frac{1}{12}\langle \Dc_{W}D_{[W^2]}\Phi\rangle+\frac{1}{6}\langle \Dc_{[W^2]}\Dc_{W}\Phi\rangle+\frac{1}{6}\langle(\Dc_{W})^3\Phi\rangle+\mathcal{O}(\tau).\\
    \end{aligned}
\end{equation}
These match \eqref{eq:U-field-first-coefficients} and serve as a check for the first few orders. Now we assume that
\begin{equation}
    I_n[\Phi]=\langle U_n\Phi\rangle+\mathcal{O}(\tau),
\end{equation}
and we will prove that this $U_n$ satisfies the recursion relation \eqref{eq:U-field-recursion}. Using \eqref{eq:Inrec2}, we have 
\begin{equation}
    n\langle U_n \Phi\rangle=\langle U_{n-1}\Dc_W \Phi\rangle+\frac{1}{2}\delta_W\langle U_{n-1} \Phi\rangle
\end{equation}
where we have isolated only $\mathcal{O}(\tau^0)$ terms. Multiplying both sides by $x^{n-1}$ and summing over $n$, we have 
\begin{equation}
\label{eq:t1}
     \langle U^{\prime}(x)\Phi\rangle=\langle U(x)\Dc_W \Phi\rangle+\frac{1}{2}\delta_W\langle U(x) \Phi\rangle
\end{equation}
Now, we use the definition of $\mathcal{W}(x)$ given in \eqref{calWdefn} and write
\begin{equation}
     {\cal W} = \sum_{n=0}^{\infty}\frac{1}{2^nn!} x^n\, [W^{n+1}]
       =e^{\frac{x\delta_W}{2}}W
\end{equation}
and we have 
\begin{equation}
\label{eq:t2}
    \Dc_{\mathcal{W}(x)}=e^{\frac{x\delta_W}{2}}\Dc_W,~~~\Dc^{\prime}_{\mathcal{W}(x)}=\frac{1}{2}\delta_{W} \Dc_{\mathcal{W}(x)}
\end{equation}
The following expressions are understood inside the trace,
which we suppress for brevity. To show that $U_n$ satisfies the recursion relation \eqref{eq:U-field-recursion}, we have to show 
\begin{equation}
    E(x)=U^{\prime}(x)-\Dc_{\mathcal{W}(x)}U(x)=0
\end{equation}
Taking $x=0$, we have $E(0)=0$. Now we take the derivative with respect to $x$ and get
\begin{equation}
\begin{aligned}
E^{\prime}(x)&=U^{\prime\prime}(x)-\Dc^{\prime}_{\mathcal{W}(x)}U(x)  - \Dc_{\mathcal{W}(x)}U^{\prime}(x) 
\end{aligned}
\end{equation}
Using \eqref{eq:t1} and \eqref{eq:t2}, we can therefore show
\begin{equation}
 E^{\prime}(x)=\frac{1}{2}\delta_W E(x)+E(x)\Dc_W   
\end{equation}
Substituting $E(x)=\sum_{n\ge 0}E_n x^n$, we get
\begin{equation}
    (n+1)E_{n+1}=\frac{1}{2}\delta_W E_n +E_n \Dc_W
\end{equation}
Since $E(0)=0=E_0$, the above recursion implies $E_n=0$ for all $n\ge0$. Thus we have
\begin{equation}
    U^{\prime}(x)=\Dc_{\mathcal{W}(x)}U(x)
    \end{equation}
    which is the operator equation~\eqref{eq:U-field-evolution}. This establishes~\eqref{eq:conjequiv2} to all orders in $x$.
Restoring the sum over sectors, the contribution with no
explicit power of $\tau$, apart from the overall factor
$\tau^{h_\Phi}$, is
$\tau^{h_\Phi}\sum_r S_{sr}
\langle\mathcal U(x)\Phi\rangle_{\tau,r}$,
where $\mathcal U(x)$ satisfies~\eqref{eq:U-field-evolution}.
\section{Summary and Outlook}
In this work, we extend the study of modular properties
of the GGE from partition functions to correlation
functions. By analysing the free-fermion two-point
function in a KdV-deformed GGE and its modular
transformation as an asymptotic series in the chemical potential, we formulate a conjecture for rational chiral CFTs. The conjecture expresses the modular transformation of an $n$-point function as a sum over sectors of correlators in the transformed ensemble, with the external fields
dressed by an operator $\mathcal U(x)$ and the ensemble deformed by the effective charge $\mathcal W(x)_0$. Both $\mathcal U(x)$ and $\mathcal W(x)$ are determined by nested OPEs involving the original current $W$.

We establish the conjecture at the lowest explicit power of the modular parameter to all orders in the chemical potential, thereby determining the dressing operator. With this operator fixed, the remaining terms provide an independent, non-trivial test of the conjecture. We verify these terms through order $\mu^3$, retaining all explicit powers of the modular parameter, and find
agreement with the proposed transformation. The main tool used in our analysis is the Zhu recursion relation satisfied by torus correlators.

Several directions remain open. The most pressing technical one is a complete proof of the factorised transformation at arbitrary order, including all explicit powers of $\tau$. The part without explicit powers of $\tau$ isolates the action of the dressing operator on the external fields, while the higher powers of $\tau$ involve the zero modes of the effective current. Controlling these requires Zhu's recursion in the presence of several zero-mode insertions.

A natural extension is to consider several chiral deformations simultaneously, following~\cite{Ashok:2026cyi}. Let
\begin{equation}
\label{eq:neww}
W(u)=\sum_{\ell}\mu_\ell I^\ell(u)\,,
\end{equation}
where the $I^\ell$ are quasi-primary currents of conformal weights $w_\ell$ and $\mu_\ell$ are their chemical potentials, and let $\lambda$ be a bookkeeping parameter that counts the total order in the deformation and is set to one at the end. Under the modular $S$ transformation,
\begin{equation}
S:~\tau\rightarrow-\frac1\tau\,,\qquad\mu_\ell\rightarrow\frac{\mu_\ell}{\tau^{w_\ell}}\,,\qquad u\rightarrow\frac u\tau\,,
\end{equation}
we expect the $N$-point function to transform as
\begin{equation}
S\Big[\big\langle\Phi\,e^{\lambda W_0}\big\rangle_{\tau,s}\Big]\sim\tau^{h_\Phi}\sum_rS_{sr}\Big\langle e^{\lambda\mathcal W(x)_0}\,\mathcal U(x)\,\Phi\Big\rangle_{\tau,r}\,,\qquad x=\frac{\lambda}{2\pi i\tau}\,,
\end{equation}
where $h_\Phi=\sum_{a=1}^Nh_a$, $S_{sr}$ is the modular $S$ matrix of the characters, and $\mathcal W(x)$ and $\mathcal U(x)$ are built from the current $W$ of~\eqref{eq:neww} exactly as before. Another extension is to full, non-chiral CFTs, where chemical potentials can be turned on for both chiralities and the left- and right-moving transformations have to be combined.

Beyond these extensions, our results open up several directions to explore. First, the transformation controls the high-temperature behaviour of correlators in GGEs, since it relates it to the low-temperature regime where the vacuum dominates. As the answer is fixed by OPE data, this behaviour is universal for a given chiral algebra. This should allow one to extend the comparisons with probes of higher-spin black holes mentioned in the Introduction to all orders in the chemical potential. This, in turn, should allow one to understand how the stationary state of an integrable CFT after a quench is probed by local operators beyond the partition function. Second, it would be interesting to understand the transformation from the point of view of the defect interpretation of GGEs~\cite{Downing:2023lnp}. In the crossed channel, the ensemble is described by a line defect. One would expect the propagation along the defect line to be governed by the effective current $\mathcal W(x)$, and the passage through an insertion point to induce the action of the dressing operator $\mathcal U(x)$. Deriving this directly from Zhu's recursion would explain why the transformation factorises, and would give a way to compute correlators in the presence of such defects. Finally, since the dressing and the effective charge are determined by nested OPEs, our results may be a useful starting point for the integrable structure underlying GGEs, for instance through the ODE/IM correspondence~\cite{Ashok:2026mfy}, where correlators have not yet been explored.

\section*{Acknowledgements}
The author thanks Diptarka Das and Sachin Grover for valuable discussions during a closely related project that inspired the present work. The author is especially grateful to Sujay K. Ashok for many fruitful discussions and helpful feedback. The author also thanks Aradhita Chattopadhyaya, Max Downing, Anirban Ganguly, Alok Laddha, Shashank Sengar, Adarsh Sudhakar, Ronak M. Soni, and Gerard M. T. Watts for valuable discussions. The author is grateful to Sujay K. Ashok, Diptarka Das, and Gerard M. T. Watts for their comments on the draft. This work is partially supported by a grant to CMI from the Infosys Foundation. 

During the preparation of this manuscript, the author used ChatGPT and Claude to assist with language editing, improve the clarity of the presentation, and check intermediate calculations. The author reviewed the AI-assisted material and takes full responsibility for the final manuscript.
\appendix
\section{Modular properties of derivatives of modular forms and Jacobi forms}
\label{app:A}
In this section, we discuss the modular properties of derivatives of modular and Jacobi forms. We begin by defining the modular transformation. Under modular transformation, the complex structure $\tau$ transforms as 
\begin{equation}   \tau\rightarrow\gamma(\tau)=\frac{a\tau+b}{c\tau+d}~,~~~ad-bc=1.
\end{equation}
Next, we shift $\tau\rightarrow\tau+x$, then we have 
\begin{equation}
    \begin{aligned}
        \gamma(\tau+x)&=\frac{a(\tau+x)+b}{c(\tau+x)+d}=\gamma(\tau)+\eta
    \end{aligned}
\end{equation}
where
\begin{equation}
 \eta=\frac{x}{j(\tau)(j(\tau)+cx)},~~~ j(\tau)=c\tau+d.  
\end{equation}
We can write $x$ as 
\begin{equation}
    x=\frac{\eta j(\tau)^2}{1-c \eta j(\tau)}.
\end{equation}
Let $f(\tau)$ be a modular form of weight $w$, then
\begin{equation}
\label{eq:defmodform}
    f(\gamma(\tau))=(c\tau+d)^w f(\tau).
\end{equation}
Under the shift $\tau\rightarrow \tau+x$, we can write \eqref{eq:defmodform} as
\begin{equation}
   \begin{aligned}
   \label{eq:mtmodshift}
        f(\gamma(\tau+x))=&(c(\tau+x)+d)^wf(\tau+x)\\
    f(\gamma(\tau)+\eta)&=\Bigg(\frac{ j(\tau)}{1-c\eta j(\tau)}\Bigg)^wf(\tau+x).
   \end{aligned}
\end{equation}
We expand the LHS of \eqref{eq:mtmodshift} and get
\begin{equation}
\label{eq:mtmodshiftlhs}
f(\gamma(\tau)+\eta)=\sum_{n=0}\frac{\eta^n}{n!}f^{(n)}(\gamma(\tau)).
\end{equation}
Next, we expand the RHS of \eqref{eq:mtmodshift} and get
\begin{equation}
\label{eq:mtmodshiftrhs}
    \begin{aligned}
        \Bigg(\frac{ j(\tau)}{1-c\eta j(\tau)}\Bigg)^w f(\tau+x)=&\Bigg(\frac{ j(\tau)}{1-c\eta j(\tau)}\Bigg)^w\sum_{m=0}\frac{x^m}{m!}f^{(m)}(\tau)\\=&\Bigg(\frac{ j(\tau)}{1-c\eta j(\tau)}\Bigg)^w\sum_{m=0}\frac{1}{m!}f^{(m)}(\tau)\Big(\frac{\eta j(\tau)^2}{1-c\eta j(\tau)}\Big)^m\\
        =&\sum_{m=0}\frac{1}{m!}f^{(m)}(\tau)\eta^m j(\tau)^{2m+w}(1-c\eta j(\tau))^{-(m+w)}
    \end{aligned}
\end{equation}
Using the following identity 
\begin{equation}
 (1-c\eta j(\tau))^{-(m+w)}=\sum_{l}\binom{m+w+l-1}{l}  (c\eta j(\tau))^l 
\end{equation}
and comparing \eqref{eq:mtmodshiftlhs} and  \eqref{eq:mtmodshiftrhs}, we get
\begin{equation}
    f^{(n)}(\gamma(\tau))=  \sum_{m=0}^{n}\frac{n!}{m!}\binom{n+w-1}{n-m}f^{(m)}(\tau)c^{n-m} j(\tau)^{w+n+m}.
\end{equation}
Since, we are interested in the modular $S$ transform, we can choose $a=0,b=-1,c=1,d=0$ and we get $j=\tau$ 
\begin{equation}
    \begin{aligned}
      f^{(n)}(S(\tau))=  \sum_{m=0}^{n}\frac{n!}{m!}\binom{n+w-1}{n-m}f^{(m)}(\tau)\tau^{w+n+m}
    \end{aligned}
\end{equation}
Next, we consider an index-zero Jacobi form $f(u,\tau)$ of weight $w$ Under the modular transformation, we have 
\begin{equation}
    f\Big(\frac{u}{j(\tau)},\gamma(\tau)\Big)=(j(\tau))^wf(u,\tau)
\end{equation}
We apply the same method and shift
$\tau\rightarrow\tau+x$ and $u\rightarrow u+s$. We then get
\begin{equation}
    f\Big(\frac{u+s}{j(\tau+x)},\gamma(\tau+x)\Big)=(j(\tau+x))^wf(u+s,\tau+x).
\end{equation}
Therefore, we can write 
\begin{equation}
     \frac{u+s}{j(\tau+x)}= \frac{u}{j(\tau)}+\zeta\Rightarrow s= \frac{ (uc\eta+\zeta) j(\tau)}{1-c\eta j(\tau)},
\end{equation}
and 
\begin{equation}
    f\Big(\frac{u}{j(\tau)}+\zeta,\gamma(\tau)+\eta\Big)=\Bigg(\frac{j(\tau)}{1-c\eta j(\tau)}\Bigg)^wf(u+s,\tau+x).
\end{equation}
Now, we expand both sides of the above equation and collect the coefficients of $\zeta$ and $\eta$, and we get

\begin{equation}
 f^{(n,m)}\Big(\frac{u}{j(\tau)},\gamma(\tau)\Big)=  \sum_{\substack{r\ge 0,\, l\ge n \\ l+r\le n+m}} \frac{m!n!}{l!r!}\binom{l}{n} \binom{-1 + w + m + n}{n+m-l-r}f^{(l,r)}(u,\tau)u^{l-n}c^{m - r}j(\tau)^{w + m + n + r},
\end{equation}
where $f^{(n,m)}(u,\tau)=\partial^{n}_u\partial^{m}_{\tau}f(u,\tau)$. Since, we are interested in the modular $S$ transform, we can choose $a=0,b=-1,c=1,d=0$ and we get $j=\tau$ and get
\begin{equation}
  f^{(n,m)}\Big(\frac{u}{\tau},S(\tau)\Big)=  \sum_{\substack{r\ge 0,\, l\ge n \\ l+r\le n+m}}\frac{m!n!}{l!r!}\binom{l}{n} \binom{-1 + w + m + n}{n+m-l-r}u^{l-n}\tau^{w + m + n + r}f^{(l,r)}(u,\tau)
\end{equation}
\section{Identities related to Bell polynomials}
\label{app:Appendix2}

In this appendix, we collect and prove the Bell-polynomial
identities used in Section \ref{sec:section4}. All generating functions
are understood as formal power series.

The partial exponential Bell polynomials $B_{N,r}(y_1,y_2,\ldots)$ are defined by
\begin{equation}
\frac{1}{r!}
\left(
\sum_{j\geq1}y_j\frac{x^j}{j!}
\right)^r
=
\sum_{N\geq r}
B_{N,r}(y_1,y_2,\ldots)\frac{x^N}{N!}.
\label{eq:bell-definition}
\end{equation}
Equivalently,
\begin{equation}
B_{N,r}(y_1,y_2,\ldots)
=
N!
\sum_{\substack{m_1,m_2,\ldots\geq0\\
                 \sum_jm_j=r\\
                 \sum_jjm_j=N}}
\prod_{j\geq1}
\frac{1}{m_j!}
\left(\frac{y_j}{j!}\right)^{m_j}.
\label{eq:bell-explicit}
\end{equation}
We use the conventions
\begin{equation}
B_{0,0}=1,
\qquad
B_{N,0}=0\quad(N>0),
\qquad
B_{N,r}=0\quad(r>N).
\label{eq:bell-boundary}
\end{equation}

Setting $y_j=j!A_j$ in
\eqref{eq:bell-definition} immediately gives
\begin{equation}
\frac{1}{r!}
\left(\sum_{j\geq1}x^jA_j\right)^r
=
\sum_{N\geq r}
\frac{x^N}{N!}
B_{N,r}(A_1,2!A_2,\ldots).
\label{eq:bell-operator-expansion}
\end{equation}
This identity also applies to mutually commuting
operators $A_j$.
From the equivalent definition \eqref{eq:bell-explicit}, we have
\begin{equation}
B_{N,r}(ab\,y_1,ab^2y_2,\ldots)
=
a^r b^N B_{N,r}(y_1,y_2,\ldots).
\label{eq:bell-homogeneity}
\end{equation}
For our purposes, we have 
\begin{equation}
A_j=a_{j-1}\partial_{\hat\tau}\partial_u^{jd},
\qquad d=2k-2.
\label{eq:bell-Aj}
\end{equation}
Each monomial in $B_{N,r}$ contains $r$ factors of
$\partial_{\hat\tau}$ and a total of $Nd$ derivatives
with respect to $u$. Therefore,
\begin{equation}
\frac{1}{N!}B_{N,r}(A_1,2!A_2,\ldots)
=
b_{N,r}\partial_{\hat\tau}^{\,r}\partial_u^{Nd},
\qquad
b_{N,r}:=\frac{1}{N!}
B_{N,r}(a_0,2!a_1,\ldots).
\label{eq:bell-operator-homogeneity}
\end{equation}
We have used the above relation in \eqref{eq:bellhomo}.

\subsection{An important identity}

We now prove
\begin{equation}
\sum_{p=l}^{n-r}
\binom{p(2k-1)}{p-l}\,b_{n-p,r}
=
\frac{1}{r!}
\binom{n(2k-1)}{n-l-r},
\qquad l+r\leq n,
\label{eq:bell-convolution}
\end{equation}
where
\begin{equation}
a_j=
\frac{1}{j+1}
\binom{(j+1)(2k-1)}{j}.
\label{eq:bell-aj}
\end{equation}

We introduce the generating function
\begin{equation}
T(x)=\sum_{j\geq1}a_{j-1}x^j.
\label{eq:bell-T}
\end{equation}
It is the unique formal power series with $T(0)=0$
satisfying
\begin{equation}
T=x(1+T)^{2k-1}.
\label{eq:bell-T-equation}
\end{equation}
We use Lagrange inversion to isolate the $x^j$ coefficient from $T(x)$. We have \cite{Gessel_2016}
\begin{equation}
T(x)|_{x^j}
=
\frac1j(1+t)^{j(2k-1)}|_{t^{j-1}}
=
\frac1j\binom{j(2k-1)}{j-1}
=
a_{j-1}.
\label{eq:bell-T-coefficients}
\end{equation}
where the notation $f(t)|_{t^j}$ is the coefficient of $t^j$ in the power series of $f(t)$. 
The Bell-polynomial definition then implies
\begin{equation}
\sum_{N\geq0}b_{N,r}x^N
=
\frac{T(x)^r}{r!}.
\label{eq:bell-b-generating}
\end{equation}
We now require the identity
\begin{equation}
\sum_{p\geq l}
\binom{p(2k-1)}{p-l}x^p
=
\frac{T(x)^l}
{1-(2k-1)x(1+T(x))^{2k-2}},
\qquad l\geq0.
\label{eq:bell-binomial-generating}
\end{equation}
To prove it, we first set $t=T$ and use the change of variable
\begin{equation}
x=\frac{t}{(1+t)^{2k-1}},
\qquad
dx=(1+t)^{-(2k-1)}
\left(1-\frac{(2k-1)t}{1+t}\right)dt.
\label{eq:bell-residue-change}
\end{equation}
Under this substitution, the denominator becomes
\begin{equation}
1-(2k-1)x(1+T(x))^{2k-2}
=
1-\frac{(2k-1)t}{1+t}.
\label{eq:bell-denominator}
\end{equation}
We multiply the right-hand side of \eqref{eq:bell-binomial-generating} with $x^{-p-1}$ and write it as  
\begin{align}
\frac{T(x)^l x^{-p-1}}
{1-(2k-1)x(1+T(x))^{2k-2}}\,dx
&=
t^l
\left(\frac{t}{(1+t)^{2k-1}}\right)^{-p-1}
(1+t)^{-(2k-1)}\,dt
\notag\\
&=
t^{l-p-1}(1+t)^{p(2k-1)}\,dt.
\label{eq:bell-residue-cancellation}
\end{align}
To extract the $x^p$ coefficient from the RHS of $\eqref{eq:bell-binomial-generating}$, we use the residue theorem and get
\begin{align}
\frac{T(x)^l}
{1-(2k-1)x(1+T(x))^{2k-2}}\Big|_{x^p}&=
\operatorname*{Res}_{t=0}
t^{l-p-1}(1+t)^{p(2k-1)}\,dt
\notag\\
&=
(1+t)^{p(2k-1)}|_{t^{p-l}}
=
\binom{p(2k-1)}{p-l}.
\label{eq:bell-binomial-proof}
\end{align}
This proves \eqref{eq:bell-binomial-generating},
including the case $l=0$.

Finally, multiply \eqref{eq:bell-b-generating} by
\eqref{eq:bell-binomial-generating}:
\begin{align}
\left(
\sum_{p\geq l}\binom{p(2k-1)}{p-l}x^p
\right)
\left(
\sum_{N\geq0}b_{N,r}x^N
\right)
&=
\frac{1}{r!}
\frac{T(x)^{l+r}}
{1-(2k-1)x(1+T(x))^{2k-2}}
\notag\\
&=
\frac{1}{r!}
\sum_{n\geq l+r}
\binom{n(2k-1)}{n-l-r}x^n.
\label{eq:bell-convolution-generating}
\end{align}
Comparing the coefficients of $x^n$ proves
\eqref{eq:bell-convolution}.
\section{Total derivatives and the dressing operator}
\label{app:total-derivatives}

In Section~\ref{sec:section4}, we identified the
effective free-fermion current $\mathcal{W}(x)$, modulo total derivatives,
with
\begin{equation}
\mathcal W(x)
\equiv
\sum_{j\geq0}a_jx^jJ_{2k+jd},
\qquad
a_j=\frac{c(j,k)}{2^jj!},
\qquad d=2k-2.
\label{eq:W-mod-total-derivatives}
\end{equation}
Equality modulo total derivatives immediately implies
equality of zero modes. It does not, in general, imply
equality of the external-field actions $\mathcal{U}(x)$, since
\begin{equation}
\mathcal D_{\partial Z}\phi=-(Z\phi)_1.
\label{eq:first-derivative-action}
\end{equation}
However, for a second-order total derivative 
\begin{equation}
\mathcal D_{\partial^2 Z}\phi=0.
\end{equation}
We show below that the ambiguity in
\eqref{eq:W-mod-total-derivatives} is a second total
derivative and therefore does not affect either
$\mathcal W(x)_0$ or $\mathcal D_{\mathcal W(x)}$.

We work with local fermion bilinears in the flat
cylinder coordinates, using the same local
normal-ordering prescription throughout. A homogeneous
bilinear of conformal weight $h$ has the form
\begin{equation}
X(u)=
\sum_{a+b=h-1}c_{ab}
:\partial_u^a\psi(u)\,\partial_u^b\psi(u):,
\label{eq:general-fermion-bilinear}
\end{equation}
where $c_{ab}=-c_{ba}$. Fermion anti-symmetry allows us to represent it by
an antisymmetric polynomial
\begin{equation}
P_X(p,q)=\sum_{a+b=h-1}c_{ab}p^aq^b,
\qquad
P_X(q,p)=-P_X(p,q).
\label{eq:bilinear-symbol}
\end{equation}
In this notation, differentiation of the entire local
field corresponds to multiplication by $p+q$:
\begin{equation}
P_{\partial X}(p,q)=(p+q)P_X(p,q).
\label{eq:derivative-symbol}
\end{equation}
We now assume that $h$ is even and therefore introduce the following change of variables  
\begin{equation}
s=p+q,\qquad v=p-q.
\label{eq:symbol-variables}
\end{equation}
Anti-symmetry under $p\leftrightarrow q$ means that
$P_X(s,v)$ is odd in $v$ and since its total degree $h-1$
is also odd, every monomial has an even power of $s$, and
\begin{equation}
P_X(s,v)=
\sum_{r=0}^{\frac{h-2}{2}}\gamma_r
s^{2r}v^{h-1-2r}.
\label{eq:even-weight-symbol}
\end{equation}
Let $X$ be a homogeneous local fermion bilinear of even
conformal weight, and suppose that $X=\partial Y$ for another
local fermion bilinear $Y$. $P_{X}$ must therefore be divisible by
$s$. Since only even powers of $s$ occur, it is in fact
divisible by $s^2$. Hence $X=\partial^2 Z$ for a local
fermion bilinear $Z$.
\begin{equation}
X=\partial Y,\quad h\ \text{even}
\qquad\Longrightarrow\qquad
X=\partial^2 Z
\label{eq:bilinear-second-derivative}
\end{equation}
This statement applies to the bilinear sector under
consideration; it is not a statement about arbitrary
local fields in a generic CFT.

The free-fermion KdV currents are even-weight
bilinears. Moreover, the second OPE pole of two such
currents is again an even-weight bilinear. At order $x^j$,
the weight is
\begin{equation}
h_j=2k+j(2k-2)=2k+jd,
\label{eq:effective-current-weight}
\end{equation}
which is even. Applying
\eqref{eq:bilinear-second-derivative} to the
total-derivative ambiguity at each order gives
\begin{equation}
\mathcal W(x)
=
\sum_{j\geq0}a_jx^jJ_{2k+jd}
+\partial_u^2 Z(x),
\qquad
Z(x)=\sum_{j\geq0}x^jZ_j.
\label{eq:W-second-derivative}
\end{equation}
Here the derivative acts on the local field, with
$x$ held fixed.

The second derivative has a vanishing zero mode:
\begin{equation}
(\partial_u^2Z)_0
=
\frac{1}{2\pi i}
\int_0^{2\pi i}du\,\partial_u^2Z(u)
=0,
\label{eq:second-derivative-zero-mode}
\end{equation}
where we used the periodicity of the bosonic bilinear.
Therefore,
\begin{equation}
\mathcal W(x)_0
=
\sum_{j\geq0}a_jx^j(J_{2k+jd})_0.
\label{eq:W-zero-mode-unchanged}
\end{equation}
To examine the dressing operator action, we write the
singular OPE as
\begin{equation}
Z(u)\phi(v)
\sim
\sum_{r\geq1}\frac{(Z\phi)_r(v)}{(u-v)^r}.
\label{eq:Z-phi-OPE}
\end{equation}
Differentiating twice with respect to $u$ gives
\begin{equation}
(\partial_u^2Z)(u)\phi(v)
\sim
\sum_{r\geq1}
\frac{r(r+1)(Z\phi)_r(v)}{(u-v)^{r+2}}.
\label{eq:second-derivative-OPE}
\end{equation}
There are no first or second poles. Hence
\begin{equation}
(\partial^2Z\,\phi)_1
=
(\partial^2Z\,\phi)_2
=0,
\qquad
\mathcal D_{\partial^2Z}\phi=0.
\label{eq:second-derivative-D-action}
\end{equation}
It follows that
\begin{equation}
\mathcal D_{\mathcal W(x)}\phi
=
\sum_{j\geq0}a_jx^j
\mathcal D_{J_{2k+jd}}\phi.
\label{eq:W-D-action-unchanged}
\end{equation}
 Thus, replacing $\mathcal W(x)$ by the bilinear representative
in \eqref{eq:W-mod-total-derivatives} leaves both $\mathcal W(x)_0$ and
$\mathcal D_{\mathcal W(x)}$ unchanged. Since the dressing
operator is uniquely determined as a formal power series by
\begin{equation}
\frac{d\mathcal U(x)}{dx}
=
\mathcal D_{\mathcal W(x)}\mathcal U(x),
\qquad
\mathcal U(0)=1,
\label{eq:U-total-derivative-independence}
\end{equation}
the same replacement also leaves $\mathcal U(x)$ unchanged. The argument applies equally to descendants of $\phi$, so it remains valid under repeated
dressing operator actions.

\section{Zhu recursion}
\label{app:Appendix3}
In this appendix, we state Zhu recursion\cite{zhu:1990,zhu:1996} in the cylinder
conventions used in the main text and derive the integrated
identity needed in Section~\ref{sec:6.1}. For a detailed review, readers may refer to \cite{Gaberdiel:2012yb, Ashok:2024zmw}. Here we use the convention
\begin{equation}
    z=e^u,
    \qquad q=e^{2\pi i\tau},
    \qquad
    u\sim u+2\pi i\sim u+2\pi i\tau.
    \label{eq:zhu-coordinate}
\end{equation}
Zhu recursion in this convention is described in \cite{tuite}. 
For a bosonic holomorphic field $A$, the zero mode is given by
\begin{equation}
    A_0=\frac{1}{2\pi i}
    \int_{a}^{a+2\pi i}du\,A(u).
\end{equation}
For a given set of local fields $X_n(u_n)$, we then define the torus $ n$-point function as the following unnormalized trace
\begin{equation}
    \langle X_1(u_1)\dots X_n(u_n)\rangle:=\Tr( X_1(u_1)\dots X_n(u_n) q^{L_0-\frac{c}{24}}),
\end{equation}
where $q=e^{2\pi i\tau}$ and $L_0-\frac{c}{24}$ is the zero mode of the energy-momentum tensor in the cylinder. It will be useful to define the mixed correlator with $l$ zero-mode insertions of field $b$ and $n$ field insertions. We have   
\begin{equation}
    \langle b_0^l X_1(u_1)\dots X_n(u_n)\rangle:=\Tr(b_0^l X_1(u_1)\dots X_n(u_n) q^{L_0-\frac{c}{24}}),
\end{equation}
As shown in \cite{Gaberdiel:2012yb}, the above satisfies the (modified) Zhu recursion relation: 
\begin{equation}
    \begin{aligned}
        &\langle b_0^l X_1(u_1)\dots X_n(u_n)\rangle=\Tr(b_0^l (X_1)_0X_2(u_2)\dots X_n(u_n) q^{L_0-\frac{c}{24}})\\
&+\sum_{i=0}^{l}\sum_{j=2}^{n}\sum_{m=0}^{\infty}\binom l ig^{i}_{m+1}(e^{(u_j-u_1)},q)\Tr(b_0^{l-i}X_2(u_2)\dots d^i[m]X_j(u_j)\dots X_n(u_n) q^{L_0-\frac{c}{24}})
    \end{aligned}
\end{equation}
where 
\begin{equation}
     P_{m+1}(e^{u},q)
    =
    \frac{1}{m!}
    \sum_{\ell\ne 0}
    \frac{\ell^m e^{l u}}{1-q^\ell},
    \qquad m\geq0.
\end{equation}
and 
\begin{equation}
     g^i_{m+1}(e^u,q)
    =
    \frac{(m-i)!}{m!}
    (q\partial_q)^i P_{m+1-i}(e^u,q),
    \qquad m\geq i\geq0,~
\end{equation}
In the recursion, the mode denoted $d^{i}[m]$ is  defined to be 
\be
d^{i}[m]=(-1)^{i}((b[0])^{i}X_1)[m]~.
\ee
Here the action of a square bracket mode can be identified with an OPE as
\begin{equation}
    A[m-1]B=(AB)_m
\end{equation}
where $(AB)_m$ is the coefficient of $(u_1-u_2)^{-m}$ in the OPE between the fields $A(u_1)$ and $B(u_2)$. A more formal definition of square bracket modes can be found in \cite{Gaberdiel:2012yb, Ashok:2024zmw}. 

Now, let us define 
\begin{equation}
    C:=\langle X_0 X_1(u_1)\dots X_n(u_n) \rangle
\end{equation}
Now, if we commute the zero mode $X_0$ through the trace and use the cyclicity property of the trace, we get
\begin{equation}
    C:=\sum_{j=1}^{n}\langle  X_1(u_1)\dots [X_0,X_j(u_j)] \dots X_n(u_n) \rangle+ C
\end{equation}
Now, the commutator can be replaced with the action square bracket mode as\cite{Gaberdiel:2012yb}
\begin{equation}
    [X_0,X_j(u_j)]=(X[0]X_j)(u_j)
\end{equation}
and we have 
\begin{equation}
\label{eq:id1}
   \sum_{j=1}^{n}\langle  X_1(u_1)\dots (X[0]X_j)(u_j) \dots X_n(u_n) \rangle=0
\end{equation}
This proves the identity \eqref{eq:id}, which we used in Section \ref{sec:6.1}.
\subsection{Relevant integrals}
Here we list the integral identities, which we
have used throughout this work:
\begin{equation}
\label{eq:revint}
     \begin{aligned}
           \int_0^{2\pi i \tau}\frac{d u_1}{2\pi i} \mathcal{P}_1(e^{u_2-u_1})&=-\frac{1}{2}+\frac{u_2-2\pi i\tau}{2\pi i}\, ,\\
            \int_0^{2\pi i \tau}\frac{d u_1}{2\pi i} \mathcal{P}_2(e^{u_2-u_1})&=\frac{1}{2\pi i}\,,\\
              \int_0^{2\pi i \tau}\frac{d u_1}{2\pi i} \mathcal{P}_m(e^{u_2-u_1})&=0\, ,\quad \forall m> 2\,,\\
\int_0^{2\pi i\tau} \frac{dw_2}{(2\pi i)}
\int_0^{2\pi i\tau} \frac{dw_1}{(2\pi i)}\,
g^1_{m+1}(e^{u_2-w_2};\tau)
&=\\
\int_0^{2\pi i\tau} \frac{dw_2}{(2\pi i)}
\int_0^{2\pi i\tau} &\frac{dw_1}{(2\pi i)}\,
P_1(e^{w_2-w_1};\tau)
P_{m+1}(e^{u_2-w_2};\tau)\\
-\frac{\delta_{m,2}}{2(2\pi i)^2}
-\Big[-
\frac{1}{2(2\pi i)}
&+\frac{1}{(2\pi i)^2}
\int_0^{2\pi i\tau} dw_2\,
P_1(e^{u_2-w_2};\tau)
\Big]\delta_{m,1},
\qquad m\geq 1.
     \end{aligned}
 \end{equation}

\section{Perturbative checks using Zhu recursion}

\label{app:D}
In this section, we will check the conjecture perturbatively. We will compute \eqref{eq:explhs} order by order and match it with \eqref{eq:cojexprhs}. We start with
\begin{equation}
   I_n[\Phi]=\frac{1}{n!}\int_{0}^{2\pi i\tau}\prod_{i=1}^{n}dw_i\langle W(w_1)\dots W(w_n) \Phi \rangle_{\tau}   
\end{equation}
For $n=0$, we have 
\begin{equation}
     I_0[\Phi]=\langle \Phi \rangle_{\tau}  ,   
\end{equation}
which matches  with \eqref{eq:cojexprhs}. For $n=1$, we have
\begin{equation}
   I_1[\Phi]=\int_{0}^{2\pi i\tau}dw\langle W(w)\Phi \rangle_{\tau}   
\end{equation}
We use Zhu recursion to get
\begin{equation}
    \begin{aligned}
        I_1[\Phi]&= (2\pi i\tau)\langle W_0\Phi\rangle_\tau+  \int_{0}^{2\pi i\tau} dw~\sum_{m\ge0}\sum_{i=1}^{N}\mathcal{P}_{m+1}(e^{(u_i-w)})~\langle (W[m]\Phi)_i\rangle_{\tau}
    \end{aligned}
\end{equation}
We use \eqref{eq:revint} and \eqref{eq:id1} to get
\begin{equation}
    \begin{aligned}
        I_1[\Phi]&= (2\pi i\tau)\langle W_0\Phi\rangle_\tau+~\langle D_W\Phi\rangle_{\tau}
    \end{aligned}
\end{equation}
Identifying $U_1=D_W$, we match \eqref{eq:cojexprhs}. For $n=2$, we have
\begin{equation}
   I_2[\Phi]=\frac{1}{2!}\int_{0}^{2\pi i\tau}dw_2dw_1\langle W(w_1)W(w_2)\Phi \rangle_{\tau}   
\end{equation}
We use Zhu recursion and get
\begin{equation}
  \begin{aligned}
    I_2[\Phi]
    =&
    \frac{2\pi i\tau}{2}
    \int_0^{2\pi i\tau}dw_2\,
    \langle W_0W(w_2)\Phi\rangle_\tau
    +\frac12
    \int_0^{2\pi i\tau}dw_2\,dw_1
    \sum_{m\geq0}
    P_{m+1}(e^{w_2-w_1})
    \langle (W[m]W)(w_2)\Phi\rangle_\tau\\
    &+\frac12
    \int_0^{2\pi i\tau}dw_2\,dw_1
    \sum_{i=1}^{N}\sum_{m\geq0}
    P_{m+1}(e^{u_i-w_1})
    \langle W(w_2)(W[m]\Phi)_i\rangle_\tau.
    \label{eq:D-I2-Zhu}
\end{aligned}  
\end{equation}
We use \eqref{eq:revint} and \eqref{eq:id1} to get
\begin{equation}
    \begin{aligned}
         I_2[\Phi]
    ={}&
    \frac{2\pi i\tau}{2}
    \int_0^{2\pi i\tau}dw\,
    \langle W_0W(w)\Phi\rangle_\tau+\frac12
    \int_0^{2\pi i\tau}dw\,
    \langle W(w)[\mathcal D_W\Phi]\rangle_\tau
    \notag\\
    &+\frac12
    \int_0^{2\pi i\tau}dw\,
    \left\langle
       \bigl[w(W[0]W)(w)+(W[1]W)(w)\bigr]\Phi
    \right\rangle_\tau.
    \label{eq:D-I2-integrated}
    \end{aligned}
\end{equation}
The second term follows directly from the $n=1$ result:
\begin{equation}
\label{eq:E8}
   \frac12
    \int_0^{2\pi i\tau}dw\,
    \langle W(w)[\mathcal D_W\Phi]\rangle_\tau
    =
    \frac{2\pi i\tau}{2}
    \langle W_0[\mathcal D_W\Phi]\rangle_\tau
    +\frac12
    \langle\mathcal D_W^2\Phi\rangle_\tau.
\end{equation}
For the last term, we use
\begin{equation}
\label{eq:E9}
  \frac12
    \int_0^{2\pi i\tau}dw\,
    \left\langle
       (W[1]W)(w)\Phi\right\rangle_\tau= \frac{2\pi i\tau}{2}
    \left\langle
      [W^2]_0\Phi
    \right\rangle_\tau+\frac12
    \langle\mathcal D_{[W^2]}\Phi\rangle_\tau. 
\end{equation}
We then evaluate the third term
\begin{equation}
    \begin{aligned}
    \label{eq:n2thirdterm}
        \frac12
    \int_0^{2\pi i\tau}dw\,w
    \left\langle
       (W[0]W)(w)\Phi
    \right\rangle_\tau= \frac12
    \int_0^{2\pi i\tau}dw\,w
  \sum_{m\ge1}\sum_{i=1}^{N}\mathcal{P}_{m+1}(e^{(u_i-w)})  \left\langle
       ((W[0]W)[m]\Phi
    )_i\right\rangle_\tau
    \end{aligned}
\end{equation}
Since $W[0]W$ is a total derivative, its zero mode and
its square-bracket zero mode vanish. For the first term
\begin{equation}
    \begin{aligned}
    \label{eq:n2zeromode}
        \frac{2\pi i\tau}{2}
    \int_0^{2\pi i\tau}dw\,
    &\langle W_0W(w)\Phi\rangle_\tau=\frac{(2\pi i\tau)^2}{2}
    \langle W^2_0\Phi\rangle_\tau+\frac{(2\pi i\tau)}{2}
    \langle W_0D_W\Phi\rangle_\tau\\
    &-\frac{(2\pi i\tau)}{2}\int_0^{2\pi i\tau}dw\,\sum_{m\ge0}\sum_{i=1}^{N}g^{1}_{m+1}(e^{u_i-w}) \langle ((W[0]W)[m]\Phi)_i\rangle_\tau
    \end{aligned}
\end{equation}

We now combine \eqref{eq:n2thirdterm} with the last term
of \eqref{eq:n2zeromode}. Using the double-integral identity in \eqref{eq:revint}
\begin{align}
&\frac12\sum_{i=1}^{N}\sum_{m\geq1}
\left[
\int_0^{2\pi i\tau}dw\,w\,P_{m+1}(e^{u_i-w})
-
\int_0^{2\pi i\tau}dw_2\,dw_1\,
g^1_{m+1}(e^{u_i-w_2})
\right]
\langle((W[0]W)[m]\Phi)_i\rangle_\tau
\notag\\
&\qquad=
\frac14\sum_{i=1}^{N}
\langle((W[0]W)[2]\Phi)_i\rangle_\tau
+
\frac12\sum_{i=1}^{N}u_i
\langle((W[0]W)[1]\Phi)_i\rangle_\tau .
\label{eq:D12}
\end{align}
In obtaining the second line, the terms independent
of $u_i$ cancel by the first-pole Ward identity. To simplify the remaining terms, we use
\begin{equation}
W[0]W
=
\frac12\partial[W^2]
+\text{total derivatives of order at least three},
\qquad [W^2]=W[1]W.
\label{eq:D13}
\end{equation}
Since
$(\partial A)[m]=-mA[m-1]$, this implies
\begin{equation}
(W[0]W)[1]=-\frac12[W^2][0],
\qquad
(W[0]W)[2]=-[W^2][1].
\label{eq:D14}
\end{equation}
Higher total derivatives do not contribute to these
two modes. Hence \eqref{eq:D12} becomes
\begin{equation}
-\frac14\sum_{i=1}^{N}
\left\langle
\bigl(u_i[W^2][0]+[W^2][1]\bigr)\Phi
\right\rangle_\tau
=
-\frac14\langle\mathcal D_{[W^2]}\Phi_i\rangle_\tau.
\label{eq:D15}
\end{equation}
Combining this with \eqref{eq:E8},\eqref{eq:E9}, and the first two
terms of \eqref{eq:n2zeromode}, we obtain
\begin{align}
I_2[\Phi]
={}&
\frac{(2\pi i\tau)^2}{2}
\langle W_0^2\Phi\rangle_\tau
+(2\pi i\tau)
\langle W_0[\mathcal D_W\Phi]\rangle_\tau
\notag\\
&+\frac{2\pi i\tau}{2}
\langle[W^2]_0\Phi\rangle_\tau
+\left\langle
\left(
\frac12\mathcal D_W^2
+\frac14\mathcal D_{[W^2]}
\right)\Phi
\right\rangle_\tau.
\label{eq:D16}
\end{align}
Thus the resulting expression agrees with the
$x^2$ coefficient of the conjecture, with
\begin{equation}
U_2=
\frac12\mathcal D_W^2
+\frac14\mathcal D_{[W^2]}.
\label{eq:D17}
\end{equation}

\subsection{The third-order contribution}

 Here, we discuss the $ n=3$ contribution. Since the $ n=3$ calculation is very cumbersome, we will not show all the steps but the important ones. For brevity, write $L=2\pi i\tau$. All integrals below
follow the prescribed separated contours, and endpoint insertions retain their inherited regulator.

We apply the previously established second-order identity
with $W(w)\Phi$ as the external insertion. Since
\begin{equation}
    I_3[\Phi]=\frac13\int_0^Ldw\,I_2[W(w)\Phi].
\end{equation}
we obtain
\begin{align}
3I_3[\Phi]
={}&\frac{L^2}{2}\int_0^Ldw\,
       \langle W_0^2W(w)\Phi\rangle_\tau
\nonumber\\
&+L\int_0^Ldw\,
       \langle W_0[\mathcal D_W(W(w)\Phi)]\rangle_\tau
+\frac L2\int_0^Ldw\,
       \langle[W^2]_0W(w)\Phi\rangle_\tau
\nonumber\\
&+\int_0^Ldw\,
 \left\langle
 \left[\left(\frac12\mathcal D_W^2
       +\frac14\mathcal D_{[W^2]}\right)(W(w)\Phi)\right]
 \right\rangle_\tau .
\label{eq:D-n3-start}
\end{align}
The action on the product is understood by the Leibniz
rule, with explicit coordinate coefficients held fixed.\\
To keep track of total derivatives, write
\begin{equation}
    W[0]W=\partial H,\qquad
    [W^2]=2H+\partial^2A,\qquad
    Z=W[1]H-\partial(W[0]A).
    \label{eq:cubic-H-A-Z}
\end{equation}
Skew symmetry and the Jacobi identity for the OPE give
\begin{equation}
    [W^2]_0=2H_0,\qquad
    \mathcal D_{[W^2]}=2\mathcal D_H,\qquad
    W[0]H+H[0]W=\partial Z.
    \label{eq:cubic-local-identities}
\end{equation}
The most non-trivial term in \eqref{eq:D-n3-start} is the term with two zero modes. The two-zero-mode Zhu recursion reads
\begin{align}
\langle W_0^2W(w)\Phi\rangle_\tau
={}&\langle W_0^3\Phi\rangle_\tau
+\sum_{i=1}^{N}\sum_{m\geq0}P_{m+1}(e^{u_i-w},q)
  \langle W_0^2(W[m]\Phi)_i\rangle_\tau
\nonumber\\
&-2\sum_{i=1}^{N}\sum_{m\geq1}g^1_{m+1}(e^{u_i-w},q)
  \langle W_0((W[0]W)[m]\Phi)_i\rangle_\tau
\nonumber\\
&+\sum_{i=1}^{N}\sum_{m\ge1}g^2_{m+1}(e^{u_i-w},q)
  \langle((W[0](W[0]W))[m]\Phi)_i\rangle_\tau .
\label{eq:cubic-mixed-Zhu}
\end{align}
Here the sum over $i$ runs over the external insertions. Using
\begin{equation}
    (W[0]W)[m]=-mH[m-1],
\qquad
(W[0](W[0]W))[m]=-m(W[0]H)[m-1].
\end{equation}
The integral identities and ordinary Zhu recursion yield
\begin{align}
\int_0^Ldw\,\langle W_0^2W(w)\Phi\rangle_\tau
={}&L\langle W_0^3\Phi\rangle_\tau
+\langle W_0^2[\mathcal D_W\Phi]\rangle_\tau
+\langle\{W_0,H_0\}\Phi\rangle_\tau
-\langle\{W_0,H(0)\}\Phi\rangle_\tau .
\label{eq:cubic-two-zero-integral}
\end{align}
Here $H(0)=\lim_{u\to 0}H(u)$ and the curly brackets denote the anti-commutator. We have used the following integral identity 
\begin{equation}
    \int_0^Ldw\,g^2_{m+1}(e^{u-w},q)
    =
    -\frac1m\left[P_m(e^u,q)+2g^1_m(e^u,q)\right],
    \qquad m\geq1.
    \label{eq:g2-integral}
\end{equation}
For the remaining terms, we use the one-zero-mode identities. We get
\begin{align}
\int_0^Ldw\,\langle W_0W(w)\Phi\rangle_\tau
={}&L\langle W_0^2\Phi\rangle_\tau
+\langle W_0[\mathcal D_W\Phi]\rangle_\tau
+\langle(H_0-H(0))\Phi\rangle_\tau ,
\label{eq:cubic-one-zero-integrals1}
\end{align}
and 
\begin{align}
    \int_0^Ldw\,
\bigl(\langle W_0H(w)\Phi\rangle_\tau
+\langle H_0W(w)\Phi\rangle_\tau\bigr)
&=L\langle\{W_0,H_0\}\Phi\rangle_\tau
+\langle W_0[\mathcal D_H\Phi]\rangle_\tau
\nonumber\\
&+\langle H_0[\mathcal D_W\Phi]\rangle_\tau
+\langle(Z_0-Z(0))\Phi\rangle_\tau .
\label{eq:cubic-one-zero-integrals2}
\end{align}
These follow from the $g^1$ integral and \eqref{eq:cubic-local-identities}. Let us discuss the boundary terms. The boundary term arises from two terms: first, from the mixed correlators of zero modes and fields. The boundary term arises from the integral $g^i$. Another way we will get a boundary term is from the total derivative. In particular,
since $\mathcal D_WW=w\partial H+2H+\partial^2A$, the term
$L\int_0^Ldw\,\langle W_0(\mathcal D_WW)(w)\Phi\rangle_\tau$
produces the endpoint contribution
$L^2\langle W_0H(L)\Phi\rangle_\tau$. Here, we use the following identity 
\begin{equation}
    \langle W_0H(L)\Phi\rangle_\tau
    =
    \langle W_0H(0)\Phi\rangle_\tau
    -\langle(W[0]H)(0)\Phi\rangle_\tau ,
\end{equation}
which we find using locality, translation and cyclicity of the trace.
The same endpoint relation applied to $\partial A$ gives
\begin{equation}
\int_0^Ldw\,
\langle W_0\partial^2A(w)\Phi\rangle_\tau
=
-\langle\partial(W[0]A)(0)\Phi\rangle_\tau.
\end{equation}
 The terms
proportional to $H(0)[\mathcal D_W\Phi]$ cancel between
the one-zero-mode integral and the cross term in
$\frac12\mathcal D_W^2(W(w)\Phi)$, which contribute
$-L\langle H(0)[\mathcal D_W\Phi]\rangle_\tau$ and
$+L\langle H(0)[\mathcal D_W\Phi]\rangle_\tau$, respectively.
The terms in which the dressing operators act entirely
on the remaining current give
\begin{align}
&\int_0^Ldw\,
\left\langle
\left[
\frac12\mathcal D_W^2W
+\frac14\mathcal D_{[W^2]}W
\right](w)\Phi
\right\rangle_\tau
\nonumber\\
&\quad=
\frac{L^2}{2}\langle(W[0]H)(0)\Phi\rangle_\tau
+L\langle(W[1]H)(0)\Phi\rangle_\tau
+\frac L2\langle Z_0\Phi\rangle_\tau
+\frac12\langle\mathcal D_Z\Phi\rangle_\tau .
\end{align}
Combining all remaining endpoint contributions to
$3I_3[\Phi]$ gives
\begin{align}
&\frac{L^2}{2}
\left\langle
\bigl([W_0,H(0)]-(W[0]H)(0)\bigr)\Phi
\right\rangle_\tau
+L\left\langle
\bigl(W[1]H-\partial(W[0]A)-Z\bigr)(0)\Phi
\right\rangle_\tau=0.
\end{align}
The coefficient of $L^2$ vanishes due to $[W_0,H(0)]=(W[0]H)(0)$, and the coefficient of $L$ by the definition
of $Z$. Thus, all endpoint terms cancel after the
contributions are combined. 
To express the result in terms of the recursively defined
currents, we use
\begin{equation}
[W^3]=[W^2][1]W+W[1][W^2]
     =4Z+2\partial(W[0]A)+\partial^2(\cdots).
\end{equation}
Consequently,
\begin{equation}
[W^3]_0=4Z_0,
\qquad
\langle\mathcal D_{[W^3]}\Phi\rangle_\tau
=4\langle\mathcal D_Z\Phi\rangle_\tau.
\end{equation}
The second derivatives have vanishing external-field action,
while the first derivative contributes
\[
-2\sum_{i=1}^{N}
\left\langle\bigl((W[0]A)[0]\Phi\bigr)_i\right\rangle_\tau=0
\]
by the Ward identity.

Collecting the surviving terms gives
\begin{align}
I_3[\Phi]
={}&\frac{L^3}{6}\langle W_0^3\Phi\rangle_\tau
+\frac{L^2}{2}\langle W_0^2[\mathcal D_W\Phi]\rangle_\tau
+\frac{L^2}{4}
 \langle\{W_0,[W^2]_0\}\Phi\rangle_\tau
\nonumber\\
&+L\left[
 \frac12\langle W_0[\mathcal D_W^2\Phi]\rangle_\tau
+\frac14\langle W_0[\mathcal D_{[W^2]}\Phi]\rangle_\tau
+\frac12\langle[W^2]_0[\mathcal D_W\Phi]\rangle_\tau
+\frac18\langle[W^3]_0\Phi\rangle_\tau
\right]
\nonumber\\
&+\left\langle
 \left[
 \frac16\mathcal D_W^3
+\frac1{12}\mathcal D_W\mathcal D_{[W^2]}
+\frac16\mathcal D_{[W^2]}\mathcal D_W
+\frac1{24}\mathcal D_{[W^3]}
 \right]\Phi
 \right\rangle_\tau .
\label{eq:cubic-final}
\end{align}
With $L=2\pi i\tau$, this agrees with the cubic coefficient
of the conjecture.

\end{document}